\documentclass[12pt,letter]{article}

\usepackage{graphicx, epsfig, color}
\def\to{\rightarrow}

\def\bi{\begin{itemize}}
\def\ei{\end{itemize}}
\def\te{\tilde e}

\def\tu{\tilde u}
\def\sps1ap{SPS1a$^\prime$}
\def\c1p{C1$^\prime$}

\def\tb{\tilde b}

\def\td{\tilde d}

\def\tst{\tilde t}
\def\ttau{\tilde \tau}

\def\tg{\tilde g}
\def\tnu{\tilde\nu}

\def\tq{\tilde q}

\def\tw{\widetilde W}
\def\tz{\widetilde Z}
\def\alt{\stackrel{<}{\sim}}
\def\agt{\stackrel{>}{\sim}}
\def\be{\begin{equation}}  
\def\ee{\end{equation}}  
\def\bea{\begin{eqnarray}}  
\def\eea{\end{eqnarray}}  
\def\beas{\begin{eqnarray*}}  
\def\eeas{\end{eqnarray*}}  
\newcommand\prd[3]{{\it Phys.\ Rev.\ }{\bf D #1} (#2) #3}

\newcommand\plb[3]{{\it Phys.\ Lett.\ }{\bf B #1} (#2) #3}

\begin{document}
\begin{titlepage}

\vspace{0.5cm}
\begin{center}
{\Large \bf Generalized focus point and mass spectra comparison \\
of highly natural SUGRA GUT models
}\\ 
\vspace{1.2cm} \renewcommand{\thefootnote}{\fnsymbol{footnote}}
{\large Howard Baer$^1$\footnote[1]{Email: baer@nhn.ou.edu }, 
Vernon Barger$^2$\footnote[2]{Email: barger@pheno.wisc.edu },
and Michael Savoy$^1$\footnote[3]{Email: savoy@nhn.ou.edu }
}\\ 
\vspace{1.2cm} \renewcommand{\thefootnote}{\arabic{footnote}}
{\it 
$^1$Dept. of Physics and Astronomy,
University of Oklahoma, Norman, OK 73019, USA \\
}
{\it 
$^2$Dept. of Physics,
University of Wisconsin, Madison, WI 53706, USA \\
}

\end{center}

\vspace{0.5cm}
\begin{abstract}
Supergravity grand unified models (SUGRA GUTs) are highly motivated and allow for
a high degree of electroweak naturalness when the superpotential parameter 
$\mu\sim 100-300$ GeV (preferring values closer to 100 GeV). 
We first illustrate that models with radiatively-driven naturalness
enjoy a generalized focus-point behavior wherein {\it all} soft terms
are correlated instead of just scalar masses.
Next, we generate spectra from four SUGRA GUT archetypes: 
1. $SO(10)$ models where the Higgs doublets live in different 10-dimensional 
irreducible representations (irreps), 
2. models based on $SO(10)$ where the Higgs multiplets live in a single 10-dimensional irrep
but with $D$-term scalar mass splitting,
3. models based on $SU(5)$ and 
4. a more general SUGRA model with 12 independent parameters.
Electroweak naturalness implies for all models a spectrum of light higgsinos 
with $m_{\tw_1,\tz_{1,2}}\alt 300$ GeV and gluinos with $m_{\tg}\alt 2-4$ TeV. 
However, masses and mixing in the third generation sfermion sector 
differ distinctly between the models. 
These latter differences would be most easily tested at a linear $e^+e^-$ collider 
with $\sqrt{s}\sim$ multi-TeV-scale but measurements at a 50-100 TeV hadron collider are also possible.
\noindent 
\vspace*{0.8cm}

\end{abstract}

\end{titlepage}

\section{Introduction}
\label{sec:intro}

Grand unified theories (GUTs) based on the gauge groups $SU(5)$\cite{su5} and $SO(10)$\cite{so10} 
present an impressive picture of both gauge group unification and matter unification and predict the
quantization of electric charge.
However, the problem of gauge hierarchy stabilization in GUT theories was noted early on.
The gauge hierarchy problem was solved via the introduction of supersymmetry\cite{gaugehierarchy} 
(SUSY) into the overall construct\cite{DG}. 
SUSY added the additional unification of fermi- and bose- degrees of freedom and received some impressive 
experimental support from the measured strength of gauge forces at LEP which were found to unify 
under renormalization group (RG) evolution within the MSSM but not within the SM\cite{gauge}. 
SUSY is also supported by the recently discovered Higgs scalar with 
$m_h\simeq 125$ GeV\cite{atlas_h,cms_h} which falls squarely within the predicted MSSM window\cite{mhiggs,h125}. 
Unification within {\it local SUSY} or supergravity 
grand unification\cite{sugra} brought gravity into the picture and offered new successes such as
a mechanism for uplifting  of the soft SUSY breaking terms. In SUGRA models, also known as gravity mediated SUSY
breaking, local SUSY is broken in a hidden sector via the superHiggs mechanism\cite{superhiggs}: 
the gravitino field absorbs the would-be Goldstino leading to a massive gravitino with value $m_{3/2}$. 
For a well-defined hidden sector, the various MSSM soft breaking terms are then all calculable as multiples of
the gravitino mass\cite{sw} which is anticipated phenomenologically to exist somewhere around the TeV scale.

This impressive construct fell into some disrepute on the experimental side via the failure to observe
flavor- and CP-violating processes, proton decay and more recently by the failure 
to discover the predicted weak scale superpartners at LHC\cite{atlas_s,cms_s}.
On the theory side, four dimensional SUSY GUTs require rather large Higgs multiplets to 
implement the GUT symmetry breaking and these seem to be inconsistent with the larger picture
where the SUGRA GUT theory might emerge from string theory\cite{bigreps}. The awkward 
role of Higgs multiplets was further exacerbated by the traditional doublet-triplet splitting problem: 
the MSSM Higgs doublets are associated with weak scale physics while the required remnant Higgs multiplets
must reside up near $Q\simeq m_{GUT}\sim 2\times 10^{16}$ GeV.

Solutions to these several Higgs-related problems were found in the formulation of
extra-dimensional GUT models. Initial models were formulated with the $SU(5)$ or $SO(10)$ GUT symmetry in 
five\cite{kawamura,altarelli,hebecker,hall5} or six\cite{hall6} spacetime dimensions. 
Orbifold compactification of the 
extra spacetime dimensions could be used as an alternative to symmetry breaking via the Higgs mechanism
as a means to break the grand unified symmetry. Such models could dispense with the large Higgs 
representations and also offer means to suppress or forbid proton decay and to solve the doublet-triplet
splitting problem\cite{altarelli}.

More recently, the rather large value of light Higgs mass $m_h\simeq 125$ GeV\cite{atlas_h,cms_h} and the lack of 
superpartners in LHC8\cite{atlas_s,cms_s} have called into question the {\it naturalness} of SUSY GUT models. 
These two disparate measurements require, in the first case, highly mixed TeV-scale top squarks 
to bolster the Higgs mass\cite{mhiggs}
and, in the second case, multi-TeV values for the gluinos and first/second generation squarks. 
Such heavy masses seem inconsistent with many calculations of upper bounds on sparticle masses
from the naturalness principle\cite{bg,dg,ac,ross} which naively requires sparticle masses 
around the 100 GeV scale. 

However, naturalness calculations using the Barbieri-Giudice (BG) measure\cite{eenz,bg} 
$\Delta_{BG}=max_i|\partial\log m_Z^2/\partial\log p_i|$ (where the $p_i$ are fundamental parameters
of the theory) were challenged\cite{comp3,seige} in that they were applied to multi-parameter
effective theories rather than the underlying SUGRA theory where all the soft terms arise as
multiples of the gravitino mass $m_{3/2}$. 
Such a misapplication of BG fine-tuning leads to {\it overestimates} of $\Delta_{BG}$ and obscures
a knowlege of which SUSY particle masses ought to lie at the 100 GeV scale.
In SUGRA theories, the appropriate parameter choices $p_i$ should be the gravitino mass $m_{3/2}$ 
and the superpotential $\mu$ parameter. Re-evaluation of $\Delta_{BG}$ in terms of 
these parameters implies that it is only the higgsinos which must lie in the 
100 GeV regime while other sparticle masses are comparable to $m_{3/2}$ which can lie 
comfortably in the multi-TeV regime\cite{ccn}: 
this latter choice is consistent with LHC8 sparticle and Higgs mass limits
and in fact was already pre-saged by the cosmological gravitino problem\cite{grav} and a decoupling solution to
the SUSY flavor and $CP$ problems\cite{dine}. 

A different naturalness measure\cite{fat,kitnom} 
$\Delta_{HS}\equiv \delta m_h^2/(m_h^2/2)$ which seemed to require several sub-TeV scale 
third generation squarks\cite{oldnsusy} was challenged as leading to overestimates of fine-tuning 
on the basis of neglecting other {\it dependent} contributions to $m_h^2$ 
which can lead to large cancellations\cite{comp3,seige,arno}. 
Regions of parameter space of the two extra-parameter non-universal Higgs model (NUHM2) were 
identified where light higgsinos $\sim 100-300$ GeV could co-exist with $m_h\sim 125$ GeV and LHC8 sparticle 
mass limits where rather mild electroweak fine-tuning at the 5-20\% level was allowed.

The question then emerges: what is the GUT basis of the NUHM2 model and are there other possibilities for 
SUGRA GUT models which allow for a high degree of EW naturalness? Some previous work was reported
which explored whether naturalness could co-exist with $b-\tau$ or $t-b-\tau$ Yukawa unified models.
To allow for $t-b-\tau$ unification, a rather large MSSM threshold correction to $m_b$ is required where\cite{pbmz,carena1,mhiggs,guasch}
\be
\Delta m_b/m_b\simeq \frac{\alpha_3\mu m_{\tg}\tan\beta}{m_{\tb}^2}+\frac{f_t^2\mu A_t\tan\beta}{m_{\tst}^2}.
\label{eq:mb}
\ee 
The required small value of $\mu$ seems to preclude Yukawa-unified natural SUSY
for $t-b-\tau$ unification better than $\sim 30\%$ and also disfavors
b-tau unification. However, it is conceivable that GUT scale
threshold corrections along with effects from compactification may evade these requirements.

In this paper, we explore several aspects of naturalness in SUGRA GUT models.
First, in Sec. \ref{sec:gfp} we show that models with radiatively-driven naturalness exhibit a 
generalized focus point behavior where weak scale contributions to $m_Z^2$ are rather insensitive to
$m_{3/2}$ for correlated choices of parameters. In Sec. \ref{sec:models} we list three SUSY GUT archetype models 
which are examined for consistency with
electroweak naturalness.\footnote{For some recent related work, see {\it e.g.} \cite{raza,todd,chin,chianese}.} 
We define these several SUGRA GUT archetype models and their associated parameter space.
These three models include: 
1. $SO(10)$ based models where the two Higgs doublets live in 
different 10-dimensional irreducible representations (irreps) (the NUHM2 model), 
2. $SO(10)$ SUSY GUT models where the two Higgs doublets live in the same 10-dimensional 
irrep (the $D$-term splitting model, DT) and 
3. a generic $SU(5)$ SUSY GUT model where $H_u\in {\bf 5}$ and $H_d\in {\bf 5^*}$. 
We will compare these results against a more general SUGRA model with 12 independent
parameters defined at the GUT scale.
In Sec. \ref{sec:nat}, we present results from a scan over each model parameter space where we identify
regions of natural SUSY parameter space. We find all four constructs allow for highly natural SUSY.
In these regions of high SUSY naturalness, we find common amongst all four models that light higgsinos
with mass $m(higgsinos) \alt 200-300$ GeV should exist and that gluinos with mass $m_{\tg}\alt 4-6$ TeV
should occur. In contrast, we find the third generation squark and slepton mixing can be very different
amongst the four models. To test such mixing, probably very high energy $e^+e^-$ colliders
with $\sqrt{s}>2m(squark,slepton)$ are needed. Some tests might be done at much higher energy
$pp$ collider with $\sqrt{s}\sim 50-100$ TeV.
A summary and conclusions are presented in Sec. \ref{sec:conclude}.

\section{Radiatively-driven naturalness as  generalized focus point behavior}
\label{sec:gfp}

To understand SUSY models with low fine-tuning, we begin with the EENZ/BG fine-tuning measure\cite{eenz,bg}
\be
\Delta_{BG}=max_i\ c_i= max_i \left|\frac{\partial\log m_Z^2}{\partial\log p_i}\right|
\ee
where the $p_i$ are fundamental parameters of the theory labeled by index $i$.
To evaluate $\Delta_{BG}$, we first express $m_Z^2$ in terms of weak scale SUSY
parameters via the well-known scalar potential minimization conditions in the MSSM 
\be
\frac{m_Z^2}{2}=\frac{m_{H_d}^2-m_{H_u}^2\tan^2\beta}{(\tan^2\beta -1)}-\mu^2\simeq -m_{H_u}^2-\mu^2
\label{eq:mzs}
\ee
where the latter partial equality holds for $\tan\beta \agt 3$.
Next, using semi-analytical solutions to the renormalization group equations for $\mu$ and $m_{H_u}^2$, 
we may express these weak scale quantities in terms of GUT scale quantities. 
It is found for example with $\tan\beta =10$ that\cite{abe,martin,feng}
\bea
m_Z^2 &=& -2.18\mu^2 +3.84 M_3^2+0.32M_3M_2+0.047 M_1M_3-0.42 M_2^2 \nonumber \\
& & +0.011 M_2M_1-0.012M_1^2-0.65 M_3A_t-0.15 M_2A_t\nonumber \\
& &-0.025M_1 A_t+0.22A_t^2+0.004 M_3A_b\nonumber \\
& &-1.27 m_{H_u}^2 -0.053 m_{H_d}^2\nonumber \\
& &+0.73 m_{Q_3}^2+0.57 m_{U_3}^2+0.049 m_{D_3}^2-0.052 m_{L_3}^2+0.053 m_{E_3}^2\nonumber \\
& &+0.051 m_{Q_2}^2-0.11 m_{U_2}^2+0.051 m_{D_2}^2-0.052 m_{L_2}^2+0.053 m_{E_2}^2\nonumber \\
& &+0.051 m_{Q_1}^2-0.11 m_{U_1}^2+0.051 m_{D_1}^2-0.052 m_{L_1}^2+0.053 m_{E_1}^2 ,
\label{eq:mHu}
\eea
where the quantities on the right-hand-side are all GUT scale parameters. 
If we evaluate the $i=Q_3$ sensitivity coefficient $\Delta_{BG}(m_{Q_3}^2)=0.73m_{Q_3}^2/m_Z^2$
and take $m_{Q_3}\agt 1$ TeV in accord with LHC sparticle limits and Higgs mass measurement, 
then we expect $\Delta_{BG}> 90$ or already about 1\% fine-tuning. 

It was observed long ago by Feng {\it et al.}\cite{fp} that if instead we assume scalar mass unification--
with $m_{H_u}=m_{Q_3}=m_{U_3} \equiv m_0$ at the GUT scale, then we can combine the contributions from lines 4 and 5 of
Eq. \ref{eq:mHu} so that $m_Z^2\sim 0.017 m_0^2$. The coefficient of the squared scalar mass term has dropped by
a factor 43: what appeared highly fine-tuned using $m_{Q_3}^2$ as a fundamental parameter is in fact low fine-tuned
when unification conditions are imposed due to cancellations between various contributions to $m_Z^2$. 
This is the {\it focus point} (FP) scenario wherein large third generation scalar
masses can be quite consistent with low fine-tuning. A related manifestation of FP SUSY is that 
for a wide range of $m_0^2$ values, then  $m_{H_u}^2$ runs to nearly the same value at $Q=m_{weak}$ 
(the RG trajectory is focused at the weak scale) so that the
value of $m_Z^2$ is relatively insensitive to variation in the high scale parameter $m_0$.

While the FP behavior reduces the fine-tuning expected in the scalar sector, there remains possible large
fine-tuning contributions to $m_Z^2$ due to the gaugino terms in Eq. \ref{eq:mHu}. 
Current limits from LHC13 imply
$m_{\tg}\simeq M_3\agt 1.5-1.8$ TeV\cite{lhc13lim}. Thus, we might expect large fine-tuning from the second term of line 1 of 
Eq. \ref{eq:mHu} as such: $\Delta_{BG}\ge c_{M_3}\agt 3.84M_3^2/m_Z^2\agt 1000$ so that SUSY appears again fine-tuned
at the $0.1\%$ level. 

At this point-- following Ref. \cite{comp3,seige}-- we recall that the soft parameters entering Eq. \ref{eq:mHu}
are only taken as independent parameters in the low energy effective theory which is expected to arise from
some more fundamental supergravity (SUGRA) or string theory. In the SUGRA theory, SUSY breaking occurs in the hidden sector
of the model and the gravitino gains a mass $m_{3/2}$ via the superHiggs mechanism\cite{sw}. 
The soft SUSY breaking terms arise from non-renormalizable terms in the SUGRA Lagrangian and are obtained by taking the
Planck mass limit $M_P\to\infty$ while keeping $m_{3/2}$ fixed. For any particular hidden sector, the soft terms are all
calculable as multiples of $m_{3/2}$ so that in reality they are all {\it dependent} terms. The soft terms are
usually taken as independent terms in the low energy effective theory only in order to parametrize the effects of
a wide range of hidden sector possibilities. By writing each soft term properly as a multiple of $m_{3/2}$ and then
combining dependent terms on the right-side of Eq. \ref{eq:mHu}, then we arrive at the simpler expression:
\be
m_Z^2\simeq -2\mu^2+ a\cdot m_{3/2}^2
\label{eq:mzsm32}
\ee
where $a$ depends on the particular spectrum which is generated. 
BG naturalness then requires $\mu^2 (GUT)\sim m_Z^2$ and $a m_{3/2}^2\sim m_Z^2$. 
Since $\mu$ hardly evolves, then equating 
$m_Z^2\simeq -2\mu^2-2m_{H_u}^2$ as a weak scale relation to Eq. \ref{eq:mzsm32} we find that 
$am_{3/2}^2\simeq m_{H_u}^2(weak)$ so that BG naturalness requires the same as tree-level 
EW naturalness\cite{rns}, namely $|m_{H_u}^2(weak)|\sim m_Z^2$.

The {\it generalized focus point} behavior is merely the observation that for certain relations amongst {\it all} the soft
parameters, a wide range of high scale input parameters $m_{H_u}^2$ can be driven to nearly the 
same weak scale values.\footnote{General conditions for focussing of $m_{H_u}^2$ at the weak scale were previously
discussed in Ref's \cite{dqm}. We thank C. Wagner for bringing these papers to our attention.}
As an example, imagine a hidden sector which produces the following soft terms:
\bea
m_0^2 &=&m_{3/2}^2\\
A_0 &=&-1.6 m_{3/2}\\
m_{1/2} &=& m_{3/2}/5\\
m_{H_d}^2 &=& m_{3/2}^2/2 .
\label{eq:soft}
\eea 
Here, we take as usual $m_0$ to be a common {\it matter scalar} soft mass which is not in general 
equal to the Higgs sector soft masses $m_{H_u}$ or $m_{H_d}$. 
We also anticipate $\mu$ to arise via some mechanism such as radiative 
PQ symmetry breaking\cite{radpq} where we take $\mu (weak)=156.6$ GeV so that $\mu (GUT)=150$ GeV.
Then, to accommodate the measured value of $m_Z=91.2$ GeV, 
we would find that the GUT scale value of $m_{H_u}^2$ is required to be
\be
m_{H_u}^2(GUT) =1.8 m_{3/2}^2-(212.52\ {\rm GeV})^2 .
\label{eq:soft2}
\ee
As we vary $m_{3/2}$ over some large range, we expect to generate values of $m_{H_u}^2(weak)$ at nearly the same values:
{\it i.e.} $|m_{H_u}^2 (weak)|$ is focused to modest values $\sim m_Z^2$ at the weak scale.

\begin{figure}[tbp]
\includegraphics[width=15cm,clip]{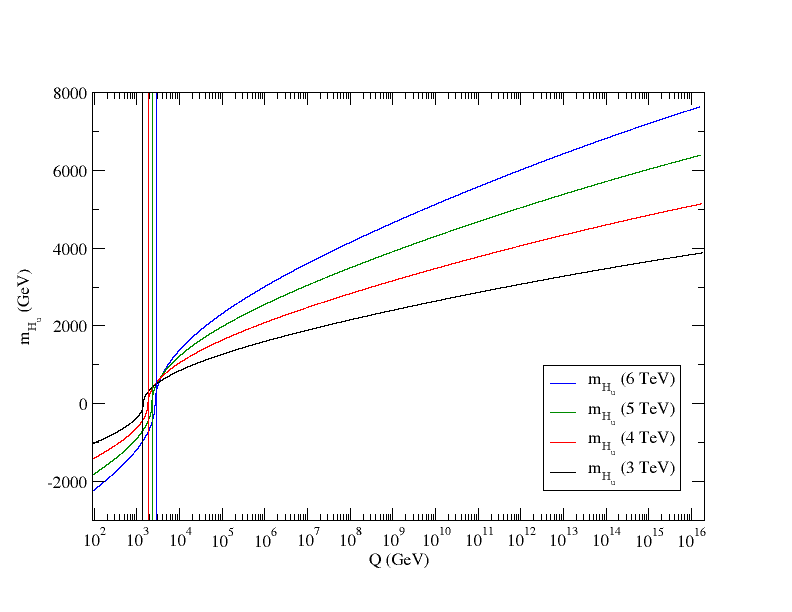}
\caption{Plot of $sign(m_{H_u}^2)\cdot \sqrt{|m_{H_u}^2|}$ vs. $Q$ 
for four different values of gravitino mass $m_{3/2}=3-6$ TeV.
\label{fig:mHu}}
\end{figure}

While the above argument makes use of the semi-analytic 1-loop RG solution for $m_Z^2$ in Eq. \ref{eq:mHu}, 
this behavior should be revealed for the usual spectrum generator codes such as Isajet and others
which make use of full 2-loop RGEs and radiatively corrected sparticle masses and scalar potential.
As an example, we show the running of $m_{H_u}^2$ versus scale $Q$ in Fig. \ref{fig:mHu} for
four choices of $m_{3/2}$: $3,\ 4,\ 5$ and 6 TeV. The locus of the $Q^2=m_{\tst_1} m_{\tst_2}$ value at which the
parameters are extracted for optimized minimization of the scalar potential are shown as vertical lines. 
We see that indeed the value of $m_{H_u}^2 (weak)$ exhibits focus point behavior for the correlated soft terms as
given in Eqs. \ref{eq:soft}-\ref{eq:soft2}.

\section{Three unified SUGRA GUT archetype models and one non-unified model}
\label{sec:models}

For all four SUSY GUT models, we assume that nature is symmetric under the GUT gauge symmetry
at energy scales $Q>m_{GUT}\simeq 2\times 10^{16}$ GeV and that below $m_{GUT}$ nature 
is described by the MSSM augmented by three gauge singlet right-hand neutrino 
superfields $N_i^c$, $i=1-3$ which are in turn integrated
out at their respective mass scales $M_{N_i}$. It is possible that
the GUT theory is a 4-dimensional quantum field theory with GUT symmetry breaking via 
Higgs multiplets\cite{pp}, or that nature is described by a $d>4$ dimensional GUT theory at $Q>m_{GUT}$ 
where the GUT symmetry is broken via compactification of the extra dimensions
via (perhaps) orbifolding\cite{kawamura,altarelli}. 
A theory of the latter type which can give rise to SUSY with radiatively-driven naturalness 
has recently been presented in Ref. \cite{cchan}.
For our numerical study, we will feign ignorance as to the GUT symmetry breaking mechanism.

\subsection{General $SO(10)$ model with each Higgs in a separate {\bf 10}: NUHM2}

For the general $SO(10)$ SUSY GUT model, we require all matter superfields to lie
in the 16-dimensional spinor representation so that matter scalar masses are unified 
to $m_{16}\ (=m_0)$.
In this model, we assume the two MSSM Higgs doublets live in different 10-dimensional
$SO(10)$ Higgs irreps so that the GUT scale Higgs soft masses $m_{H_u}^2$ and $m_{H_d}^2$ are
independent parameters. Also, in this model one might expect under the simplest conditions to have 
$b-\tau$ Yukawa coupling unification but not $t-b-\tau$ Yukawa unification. 
For ease of computing within the restrictions of natural models, 
we trade the GUT scale inputs $m_{H_u}^2$ and $m_{H_d}^2$ in lieu of weak scale parameters
$\mu$ and $m_A$. For this model, then, the relevant parameter space is that of the well-known 
two-extra-parameter non-universal Higgs mass model also known as NUHM2\cite{nuhm2}:
\be
m_0,\ m_{1/2},\ A_0,\ \tan\beta ,\ \mu ,\ m_A\ \ \ \ (NUHM2) .
\ee

We scan over the following parameters:
\bea
m_0 &:& \ 0-20\ {\rm TeV}, \nonumber\\
m_{1/2} &:& \  0.2-3\ {\rm TeV},\nonumber\\
-3 &<& A_0/m_0 \ <3,\nonumber\\
\mu &:& \ 0.1-0.5\ {\rm TeV}, \label{eq:nuhm2param}\\
m_A &:& \ 0.15-20\ {\rm TeV},\nonumber\\
\tan\beta &:& 3-60 . \nonumber
\eea

We take the various generations of scalar soft terms to be degenerate
as is suggested by the degeneracy solution to the SUSY flavor and CP problems.
We require of our solutions that:
\bi
 \item electroweak symmetry be radiatively broken (REWSB),
 \item the neutralino $\tz_1$ is the lightest MSSM particle,
 \item the light chargino mass obeys the model
independent LEP2 limit, $m_{\tw_1}>103.5$~GeV~\cite{lep2ino},
\item LHC8 search bounds on $m_{\tg}$ and $m_{\tq}$ 
from the $m_0$ vs. $m_{1/2}$ plane\cite{atlas_s} are respected,
\item $m_h=125\pm 2$~GeV.
\ei

The calculational framework allowing weak scale $\mu$ and $m_A$ inputs in lieu of $m_{H_u}^2$ 
and $m_{H_d}^2$ is encoded in Isajet/Isasugra versions $\ge 7.72$.\cite{nuhm2}. 
For the spectra calculations presented here, we use Isajet 7.85\cite{isajet}.
Here we do not enforce $b-\tau$ Yukawa coupling unification, thus allowing for GUT scale 
threshold effects which may modify this relation. 

\begin{figure}[tbp]
\includegraphics[width=15cm,clip]{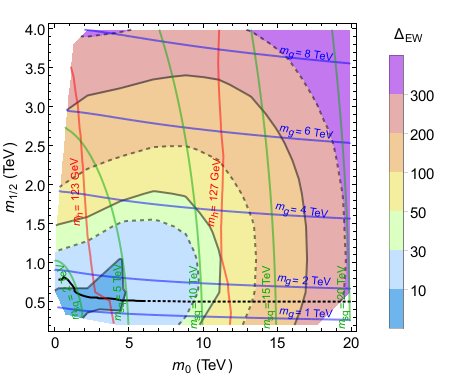}
\caption{Plot of contours of $\Delta_{EW}$ in the $m_0$ vs. 
$m_{1/2}$ plane for $\tan\beta =10$ and $A_0=-1.6 m_0$ with
$\mu =150$ GeV and $m_A=1$ TeV. We also show contours of $m_{\tg}$, 
average squark mass $m_{\tq}=2,\ 5,\ 10,\ 15$ and 20 TeV and Higgs mass $m_h$.
We show the LHC8 Atlas excluded region region below the black contour.
\label{fig:nuhm2plane}}
\end{figure}

The $m_0$ vs. $m_{1/2}$ parameter space plane of NUHM2 is shown in Fig. \ref{fig:nuhm2plane} for
$\tan\beta =10$, $A_0=-1.6 m_0$ with $\mu =150$ GeV and $m_A=1$ TeV. 
We also show contours of Higgs mass (red),
gluino mass (blue) and average first generation squark mass (green). The color-coded regions show
$\Delta_{EW}<10$ (blue) in the lower left and $\Delta_{EW}<30$ (light-blue). These highly natural regions 
can lie well beyond the current reach limits from LHC8 and also beyond the ultimate reach of LHC14 with 
300-1000 fb$^{-1}$ of integrated luminosity. 
As one moves to larger values of $m_0$ and $m_{1/2}$, 
the model becomes increasingly fine-tuned and unnatural.

\subsection{$SO(10)$ model with Higgs in a single {\bf 10}: $D$-term splitting}

For this model, we again assume that the matter superfields live in the 16-dimensional spinorial
irrep of $SO(10)$ so that matter is unified as well as forces. But now we will assume
that there is a single {\bf 10} of Higgs $\phi (10)$ (which contains both a {\bf 5} and a $\bf 5^*$) 
of $SU(5)$ Higgses and that the MSSM
Higgs doublets are both elements of the same 10-dimension GUT Higgs rep. 
We will assume in this case that the GUT scale Higgs mass splitting arises from $D$-term contributions 
to scalar masses which arise from the $SO(10)$ breaking. 
The $D$-term splitting also gives a well-defined pattern of matter scalar mass splittings
and moreover these splittings are correlated with the Higgs splitting:
\bea
m_Q^2&=&m_E^2=m_U^2=m_{16}^2+M_D^2\nonumber \\
m_D^2&=&m_L^2=m_{16}^2-3M_D^2 \nonumber \\
m_{H_{u,d}}^2&=&m_{10}^2\mp 2M_D^2 \nonumber \\
m_N^2&=&m_{16}^2+5M_D^2
\label{eq:dterms}
\eea
where $M_D^2$ parametrizes the magnitude of the  $D$-term splitting.
The value of $M_D^2$ can be taken as a free parameter of order the weak scale
owing to our ignorance of the gauge symmetry breaking mechanism.
It can take either positive or negative values.
Thus, the $DT$ model is initially characterized by the following six free parameters:
\be
m_{16},\ m_{10},\ M_D^2,\ m_{1/2},\ A_0,\ sign(\mu ),\ \tan\beta .
\ee

Here, the GUT scale soft breaking Higgs masses $m_{H_u}^2$ and $m_{H_d}^2$
are determined by $m_{10}$ and $M_D^2$. These input parameters are rather awkward for generating
SUSY models with electroweak naturalness where $\mu$ is required to be small.
The main problem is that electroweak symmetry is {\it barely broken} in radiative natural SUSY\cite{rns} 
($|m_{H_u}^2(weak)|\sim m_Z^2$) and since we use an iterative up-down running approach to the RG solution, 
EW symmetry must be properly broken on each iteration in order to generate a viable mass spectrum.
In barely-broken electroweak symmetry breaking, frequently EW symmetry will fail to be broken on some iteration
so then the whole calculation fails.

A better scheme for natural SUSY is to use $\mu$ and $m_A$ as input parameters which then
determine $m_{H_u}$ and $m_{H_d}$ at the weak scale. The values of $m_{H_u}^2$ and $m_{H_d}^2$
are then run from $m_{weak}$ to $m_{GUT}$ using the RGEs to determine their GUT scale
values. At $Q=m_{GUT}$, the required value of $m_{10}$ and $M_D^2$ can be determined
as outputs instead of inputs. 
To implement this scheme, we have programmed this new model into Isajet 7.85 as
model choice \#11: 
the NUHM $D$-term splitting model (DT). The DT model functions similarly to the NUHM2 model
except that now the matter scalars are split according to Eq. \ref{eq:dterms} at the GUT scale.
Thus, for the DT model, we will adopt the parameter space:
\be
m_0,\ m_{1/2},\ A_0,\ \tan\beta,\ \mu,\ m_A \ \ \ (DT)
\ee
where the first three are GUT scale inputs while the latter three are weak scale inputs
and where we take $m_{16}\equiv m_0$.
In this case, $M_D^2$ and $m_{10}$ are outputs of the code.
While the parameter space is the same as the NUHM2 model, the spectrum is quite different since
now there is  matter scalar splitting which is correlated with he GUT scale Higgs soft term splitting.

In this simple model, a high degree of $t-b-\tau$ Yukawa coupling unification would be expected 
in the simplest models. 
However, previous investigations find this  difficult to reconcile with
natural SUSY\cite{bkk} due to a suppression by the small $\mu$ parameter 
of the needed weak scale threshold effects.

For the $DT$ model, we will scan the same range of parameters as in the NUHM2 case.

\subsection{$SU(5)$ model}

For simplicity, we assume that the MSSM+right-hand-neutrino (RHN) model is the 
correct effective field theory below $Q=m_{GUT}$ but that the MSSM boundary conditions 
at $Q=m_{GUT}$ respect the $SU(5)$ symmetry. 
Thus, the parameter space of the model is given by
\be
m_5,\ m_{10},\ m_{1/2},\ A_t,\ A_b = A_\tau,\ \tan\beta ,\ \mu,\ m_A\ \ \ (SU(5)) 
\ee
where as usual the $L_i$ and $D_i$ superfields live in a ${\bf 5^*}$ $\psi^j$ and the $Q_i$, $U_i$ and $E_i$ 
live in a $\bf 10$ $\phi_{jk}$ of $SU(5)$. 
The index $i$ is a generation index while $j,k$ are $SU(5)$ indices.
One Higgs doublet $H_u$ lives in a ${\bf 5}$ of Higgs while the $H_d$ lives in a ${\bf 5^*}$ Higgs irrep.
Here as usual we have traded the two GUT scale Higgs doublet soft masses $m_{H_u}^2$ and $m_{H_d}^2$
in favor of the weak scale parameters $\mu$ and $m_A$. 
Since we use $\mu$ and $m_A$  as an input parameters, we may use Eq. \ref{eq:mzs} to compute 
the required weak scale values of $m_{H_u}^2$ and $m_{H_d}^2$ so as to enforce the 
measured value of $m_Z$.
The values of $m_{H_u}^2$ and $m_{H_d}^2$ are then run from $Q=m_{weak}$ to $Q=m_{GUT}$ according to
their RGEs resulting in non-universal GUT scale scalar masses. 
Since the MSSM Higgs doublets are required to occur in separate ${\bf 5}$ and ${\bf 5^*}$ reps 
of $SU(5)$, this scheme is in accord with $SU(5)$ gauge symmetry.

For our parameter space scans, we will scan the $SU(5)$ model over the following ranges:
\bea
m_{5,10} &:& \ 0.1-20\ {\rm TeV}, \nonumber\\
m_{1/2} &:& \  0.2-3\ {\rm TeV},\nonumber\\
-40 &<& A_{t,b} <40\ {\rm TeV},\nonumber\\
\mu &:& \ 0.1-0.5\ {\rm TeV}, \label{eq:su5param}\\
m_A &:& \ 0.15-20\ {\rm TeV},\nonumber\\
\tan\beta &:& 3-60 . \nonumber
\eea

\subsection{SUGRA model with 12 free parameters: SUGRA12}

For purposes of comparison, we will contrast the above results with those of a model
which includes RGE running but where the GUT scale soft scalar masses are unrelated.
This is in accord with assuming that the SM gauge symmetry is valid at $Q>m_{GUT}$
although we do still maintain gaugino mass unification (gaugino mass non-universality
for highly natural SUSY models is explored in Ref. \cite{inos}.)
We will again trade the GUT scale values of $m_{H_u}$ and $m_{H_d}$ in lieu of
weak scale values $\mu$ and $m_A$. 
This is the 12-free-parameter SUGRA model\footnote{A subset of the 19 free parameter SUGRA model\cite{sug19} where gaugino masses are unified and generations are unified.} 
with parameter space given by
\be
m_{Q,U,D,L,E},m_{1/2},A_t,A_b,A_\tau,\mu,m_A,\tan\beta \ \ \ (SUGRA12)
\ee
where we assume all three generations of matter scalars are degenerate in accord with 
a degeneracy solution to the SUSY flavor and $CP$ problems\cite{masiero,arkani}.
This model is susceptible to large contributions to unnaturalness from 
electroweak $D$-term contributions to scalar masses\cite{DTnat}.

For the SUGRA12 model, we scan over the following range:
\bea
m_{Q,U,D,L,E} &:& \ 0.1-20\ {\rm TeV}, \nonumber\\
m_{1/2} &:& \  0.2-3\ {\rm TeV},\nonumber\\
-40 &<& A_{t,b,\tau } \ <40\ {\rm TeV},\nonumber\\
\mu &:& \ 0.1-0.5\ {\rm TeV}, \label{eq:sug12}\\
m_A &:& \ 0.15-20\ {\rm TeV},\nonumber\\
\tan\beta &:& 3-60 . \nonumber
\eea

\subsection{$b$-$\tau$ Yukawa unification}

As a first examination, we compute the degree of $b-\tau$ Yukawa 
coupling unification vs. $\Delta_{EW}$ from each of the four models.
We quantify the degree of Yukawa coupling unification via
\be
R_{b\tau}=\max (f_b,f_\tau)/\min (f_b,f_\tau) .
\label{eq:Rbtau}
\ee
where the Yukawa couplings $f_b$ and $f_\tau$ are understood to be GUT scale values.

\begin{figure}[tbp]
\includegraphics[width=8cm,clip]{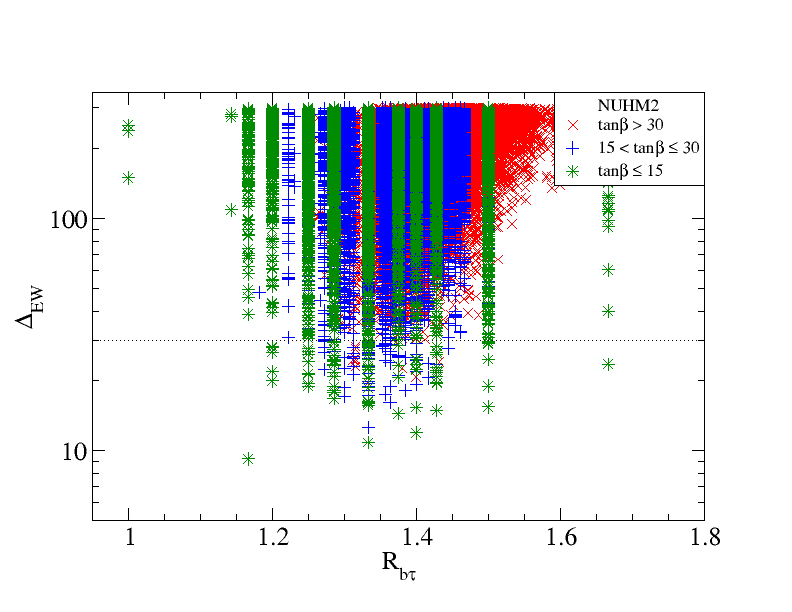}
\includegraphics[width=8cm,clip]{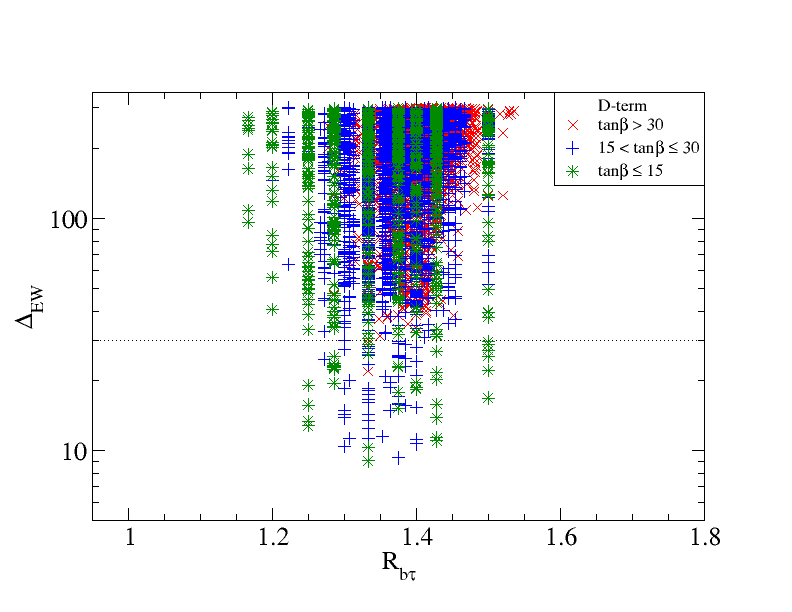}\\
\includegraphics[width=8cm,clip]{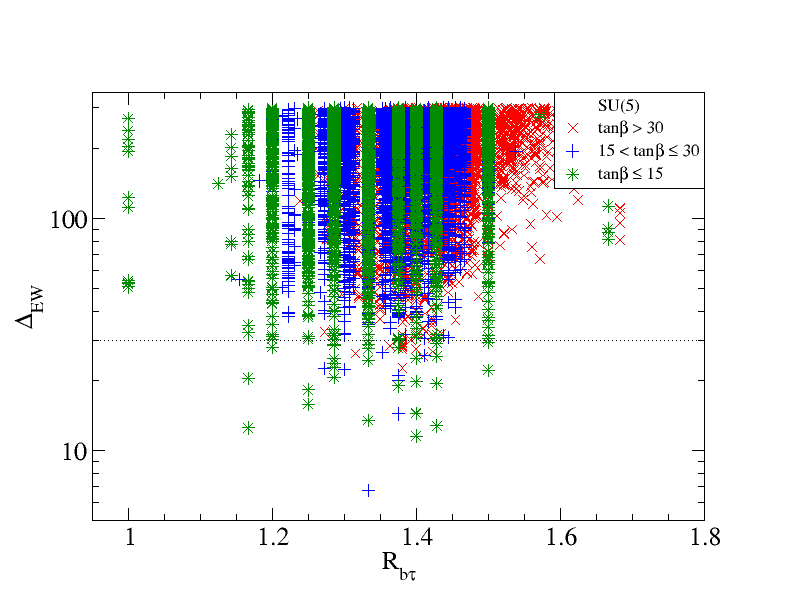}
\includegraphics[width=8cm,clip]{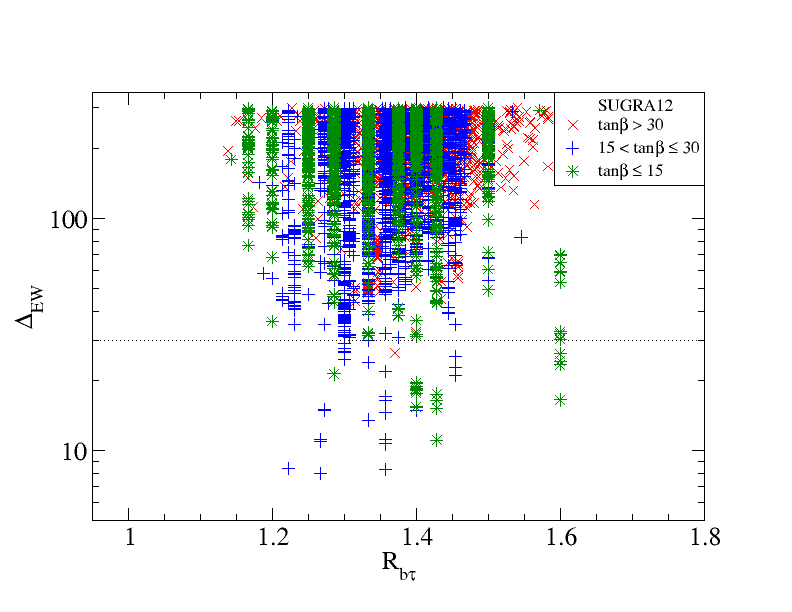}
\caption{Plot of $\Delta_{EW}$ vs. $R_{b\tau}$ for {\it a}) the NUHM2 model, 
{\it b}) the D-term model, {\it c}) the $SU(5)$ model and {\it d}) 
the SUGRA12 model. 
\label{fig:Rbtau}}
\end{figure}

In Fig. \ref{fig:Rbtau}, our results are shown for the four models
with color coded points corresponding to $\tan\beta <15$ (green), 
$15<\tan\beta  <30$ (blue) and $\tan\beta >30$ (red). 
Points with $R_{b\tau}=1$ would have exact $b-\tau$ unification at 
$Q=m_{GUT}$. 

The first point of emphasis is that low $\Delta_{EW}$ ranging as low as 10 
($\Delta_{EW}^{-1}=10\%$ electroweak fine-tuning) solutions can be found for all four
models. 
For a second point, from frame {\it a}) we see that in the NUHM2 model 
$R_{b\tau}\simeq 1$ does occur for several solutions but with rather high $\Delta_{EW}>100$.
For very natural models with $\Delta_{EW}<30$, then $b-\tau$ Yukawa couplings 
unify at the $R_{b\tau}\sim 1.2-1.5$ level.
Generally, to allow for $b-\tau$ unification, one needs a large 
one-loop $b$-quark threshold correction (see Eq. \ref{eq:mb}) but with $\mu$ small for
low $\Delta_{EW}$ solutions, this is never large.
These results appear uniform across all four models although for $SU(5)$ we did find
some $b-\tau$ unified solutions with $\Delta_{EW}$ as low as $\sim 50$.

\section{Naturalness in SUGRA GUT models: numerical results}
\label{sec:nat}

\subsection{Gluino, wino and bino masses}

\begin{figure}[tbp]
\includegraphics[width=8cm,clip]{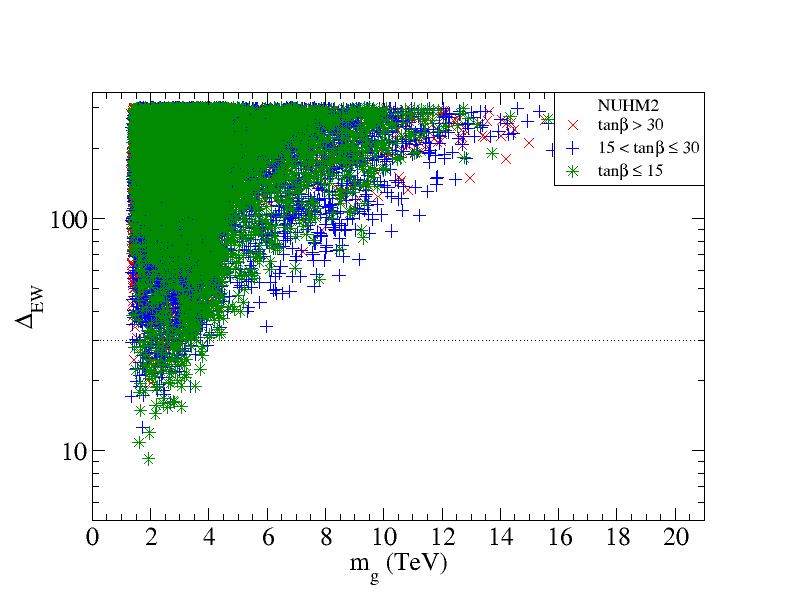}
\includegraphics[width=8cm,clip]{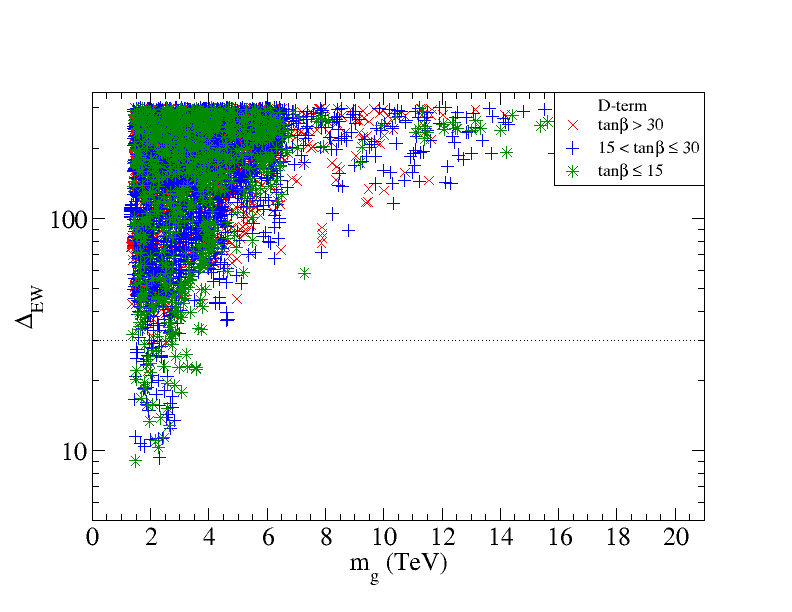}\\
\includegraphics[width=8cm,clip]{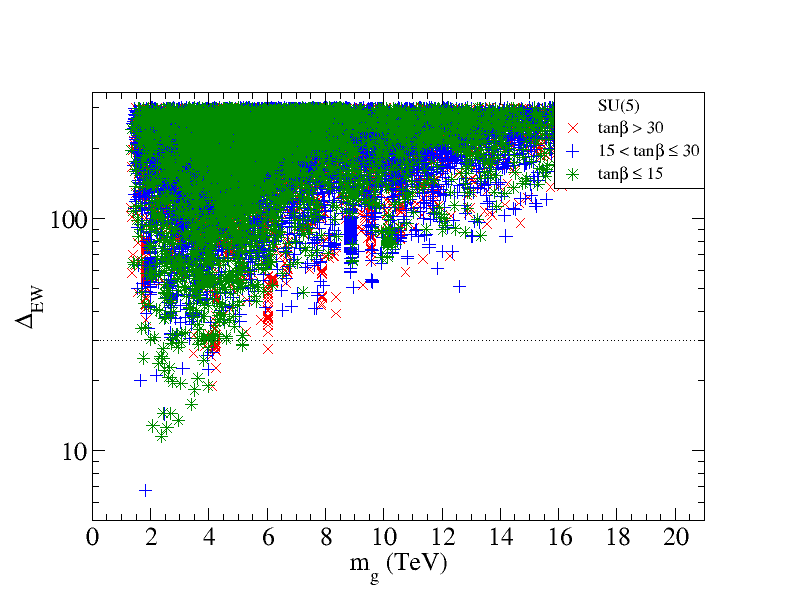}
\includegraphics[width=8cm,clip]{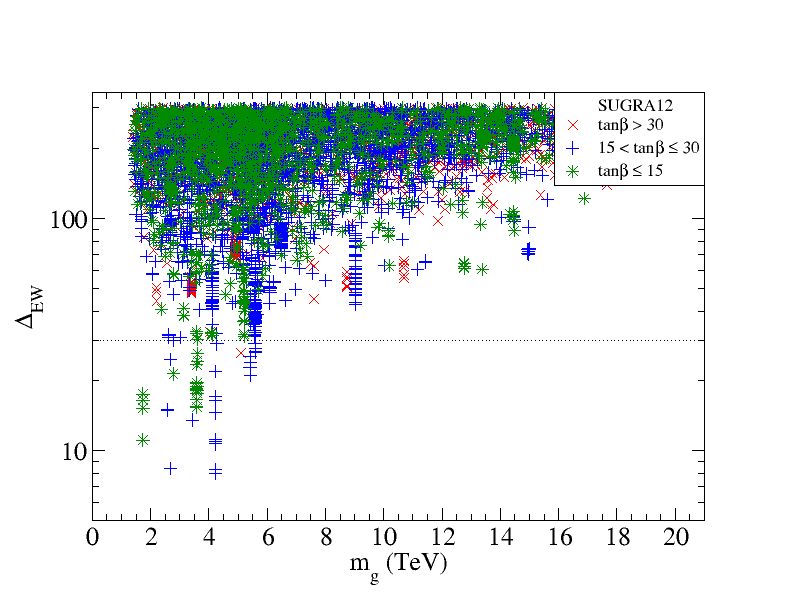}
\caption{Plot of $\Delta_{EW}$ vs. $m_{\tg}$ for {\it a}) the NUHM2 model, 
{\it b}) the DT model, {\it c}) the $SU(5)$ model and {\it d}) 
the SUGRA12 model. 
\label{fig:mgl}}
\end{figure}

For our numerical mass results from a scan over the four SUGRA GUT models, we
show in Fig. \ref{fig:mgl} the value of $m_{\tg}$ vs. $\Delta_{EW}$ 
for each case. In frame {\it a}), we find, 
as shown earlier in Ref's \cite{rns,upper} that for $\Delta_{EW}<30$ then
$m_{\tg}\alt 4$ TeV in the NUHM2 model. 
This bound arises due to the contribution of the running $SU(3)$ gaugino mass $M_3$ 
on the values of $m_{\tst_{1,2}}$; these latter values enter $\Delta_{EW}$ 
via the $\Sigma_u^u(\tst_{1,2})$ terms.
In frame {\it b}) for the DT model, the upper bound on $m_{\tg}$
is comparable if not slightly stronger: $m_{\tg}\alt 3.5$ TeV. 

In contrast, the less constrained $SU(5)$ and SUGRA12 models shown in frames {\it c}) and {\it d}) 
allow a weaker bound on $m_{\tg}\alt 6$ TeV. 
These bounds are slightly stronger than the corresponding bounds from the pMSSM model 
(with no RG running) shown in Ref. \cite{upper} where
$m_{\tg}\alt 7$ TeV due to 2-loop contributions to the scalar potential\cite{dedes}. 
In comparison with these mass bounds, we remark that the
$5\sigma$ reach of LHC14 for gluino pair production extends to about 
$m_{\tg}\sim 2$ TeV for 300-1000 fb$^{-1}$ of integrated luminosity\cite{andre}.
Thus, LHC14 will be able to probe only the lower range of $m_{\tg}$ allowed by natural SUSY.

In the models presented here, we always assume gaugino mass unification 
$M_1=M_2=M_3$ at the GUT scale. 
RG evolution then leads to $7 M_1\sim 3.5 M_2\sim M_3$
at the weak scale for the bino, wino and gluino masses respectively. 
As gaugino mass bounds for $\Delta_{EW}<30$,
we find that the bino mass $M_1\alt  600$ GeV for NUHM2 and the DT model, but 
$M_1\alt 900$ GeV for $SU(5)$ and SUGRA12. Likewise, we find that the
wino mass $M_2\alt 1200$ GeV for NUHM2 and DT models but $M_2\alt 1800$ GeV for
$SU(5)$ and SUGRA12 models.

%
%
%
%

\subsection{$\mu$ parameter}

\begin{figure}[tbp]
\includegraphics[width=8cm,clip]{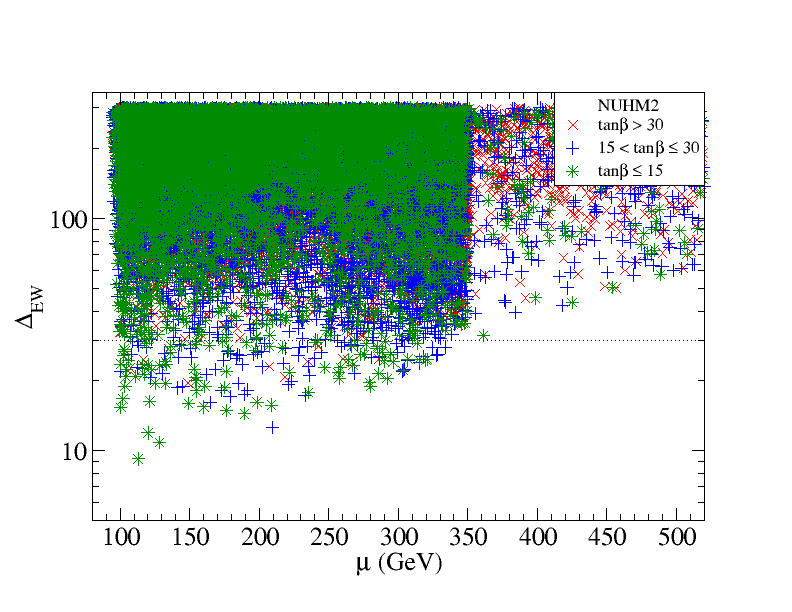}
\includegraphics[width=8cm,clip]{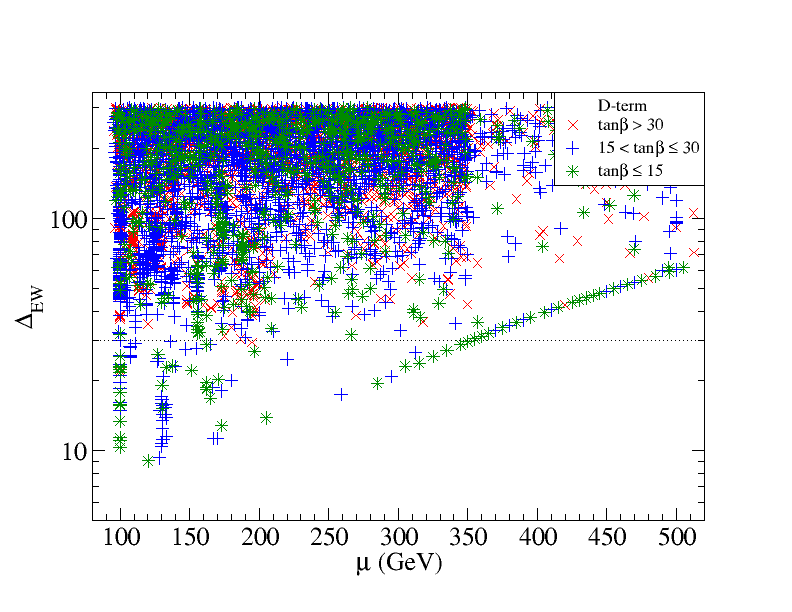}\\
\includegraphics[width=8cm,clip]{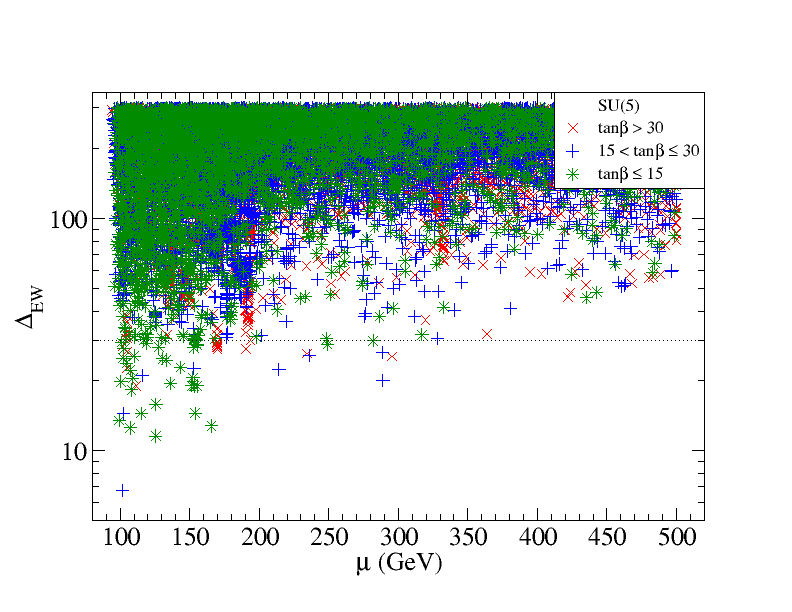}
\includegraphics[width=8cm,clip]{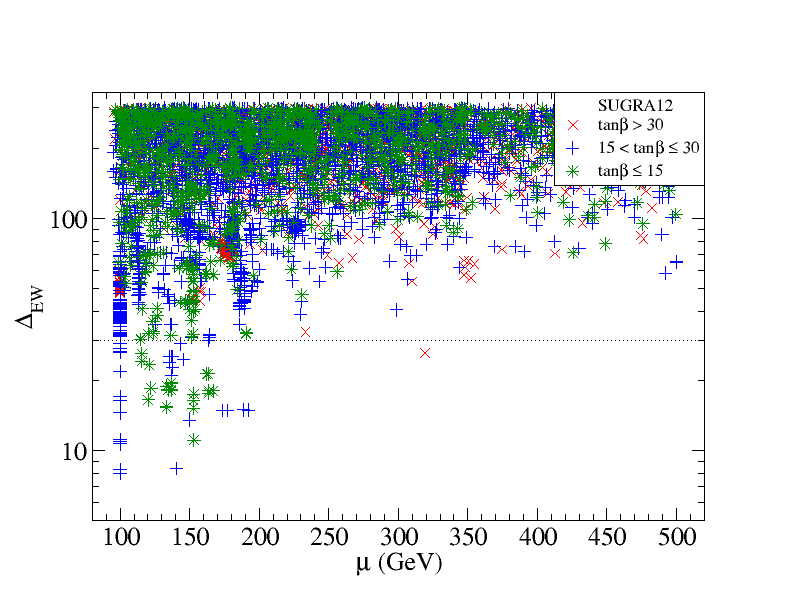}
\caption{Plot of $\Delta_{EW}$ vs. $\mu$ for {\it a}) the NUHM2 model, 
{\it b}) the D-term model, {\it c}) the $SU(5)$ model and {\it d}) 
the SUGRA12 model. 
\label{fig:mu}}
\end{figure}

The magnitude of the superpotential $\mu$ parameter is highly restricted by
Eq. \ref{eq:mzs} to lie not too far from $m_Z$ or $m_h$. 
Indeed, from Fig. \ref{fig:mu} we see that for $\Delta_{EW}<30$ then
$\mu\alt 350$ GeV for all cases since the mu parameter enters $\Delta_{EW}$
at tree level. 
This is the most robust prediction of electroweak naturalness for SUSY models.
It leads to the presence of four light higgsino-like charginos and neutralinos $\tw_1^\pm$, 
 $\tz_{1,2}$ with mass $\sim 100-350$ GeV. The mass splittings amongst
the higgsinos $m_{\tw_1}-m_{\tz_1}$ and $m_{\tz_2}-m_{\tz_1}$ are governed by
how heavy the binos and winos are, and as seen from the last section these are
also restricted by naturalness. Thus, typically from natural SUSY we obtain mass
splittings $\sim 10-30$ GeV. Tinier mass splittings require a larger 
gaugino-higgsino mass gap but this splitting cannot get arbitrarily large
according to the last subsection.
Larger mass splittings can be obtained from models with gaugino mass
non-universality\cite{inos}.
The expected small mass splittings mean that higgsino pair production
at LHC results in events with very soft tracks which are difficult to 
trigger on much less than distinguish from SM background processes.
The light higgsinos should be easily observed in the clean 
environment of an $e^+e^-$ collider with $\sqrt{s}>2m(higgsino)$\cite{ilc}.

\subsection{Third generation sfermion masses and mixing}

\begin{figure}[tbp]
\includegraphics[width=8cm,clip]{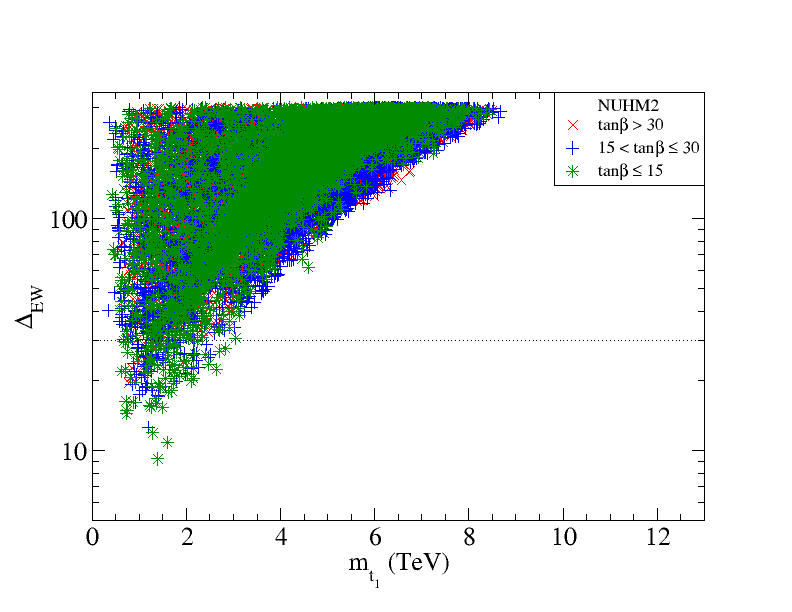}
\includegraphics[width=8cm,clip]{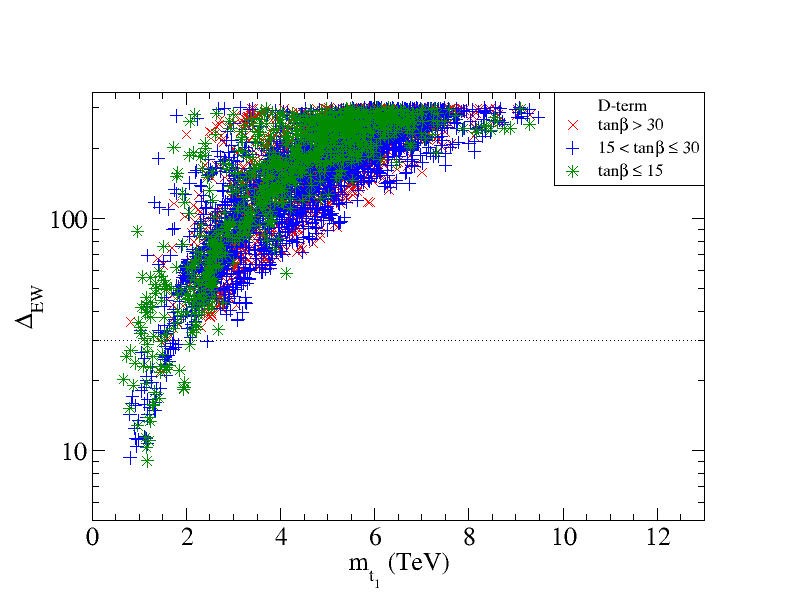}\\
\includegraphics[width=8cm,clip]{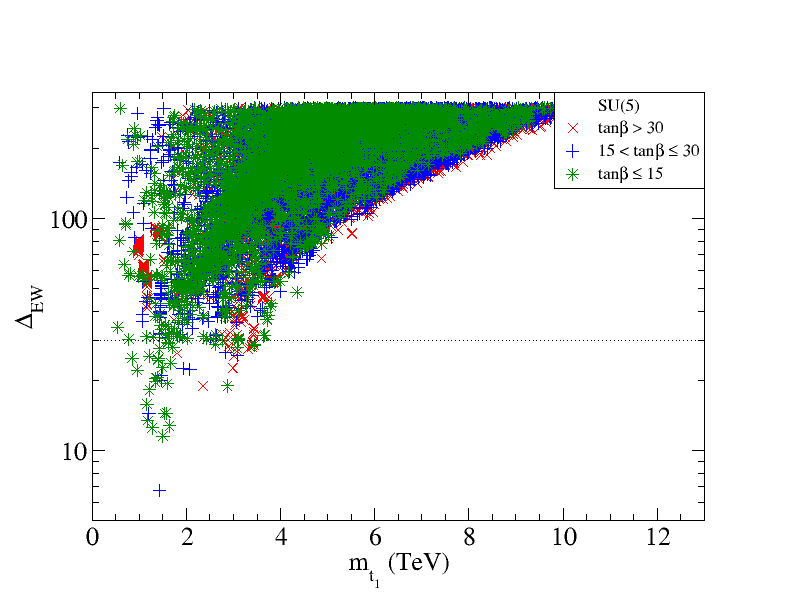}
\includegraphics[width=8cm,clip]{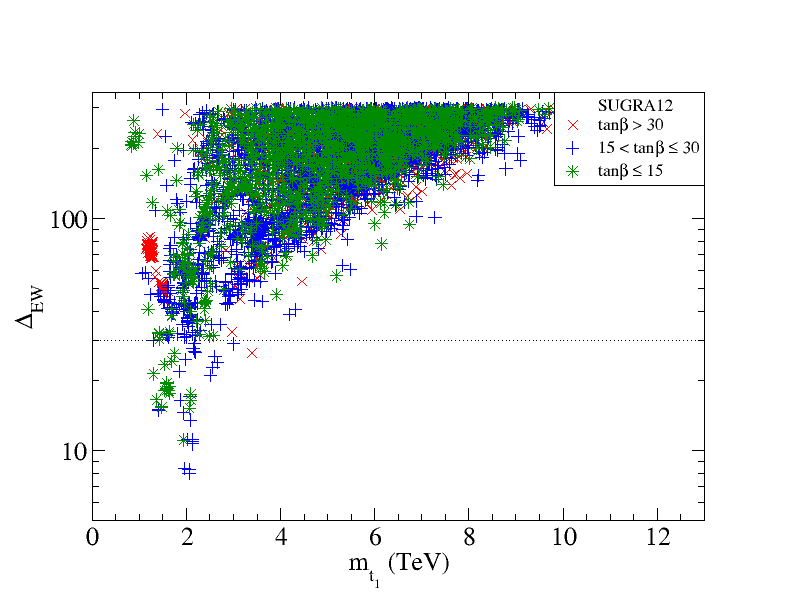}
\caption{Plot of $\Delta_{EW}$ vs. $m_{\tst_1}$ for {\it a}) the NUHM2 model, 
{\it b}) the D-term model, {\it c}) the $SU(5)$ model and {\it d}) 
the SUGRA12 model. 
\label{fig:mt1}}
\end{figure}

In Fig. \ref{fig:mt1} we show the lightest top squark mass $m_{\tst_1}$ vs.
$\Delta_{EW}$ for each of four models. The top squark masses have sharp upper
bounds due to the $\Sigma_u^u(\tst_{1,2})$ terms in Eq. \ref{eq:mzs}.
The precise contributions are listed in Ref. \cite{rns}. For the NUHM2, $SU(5)$ and SUGRA12
models we find $m_{\tst_1}\alt 3$ TeV for $\Delta_{EW}<30$. 
For the DT model, this bound seems tightened slightly to $m_{\tst_1}\alt 2$ TeV. 
These upper bounds are much higher than expected from old natural SUSY models\cite{oldnsusy} 
where three third generation squarks with mass $\alt 500$ GeV were expected. For comparison, 
the reach of LHC14 in terms of $m_{\tst_1}$ is to the 1 TeV vicinity for
various simplified models. Thus, as in the case of the gluino, 
natural SUSY can easily evade LHC stop searches with stops in the 1-3 TeV region.

\begin{figure}[tbp]
\includegraphics[width=8cm,clip]{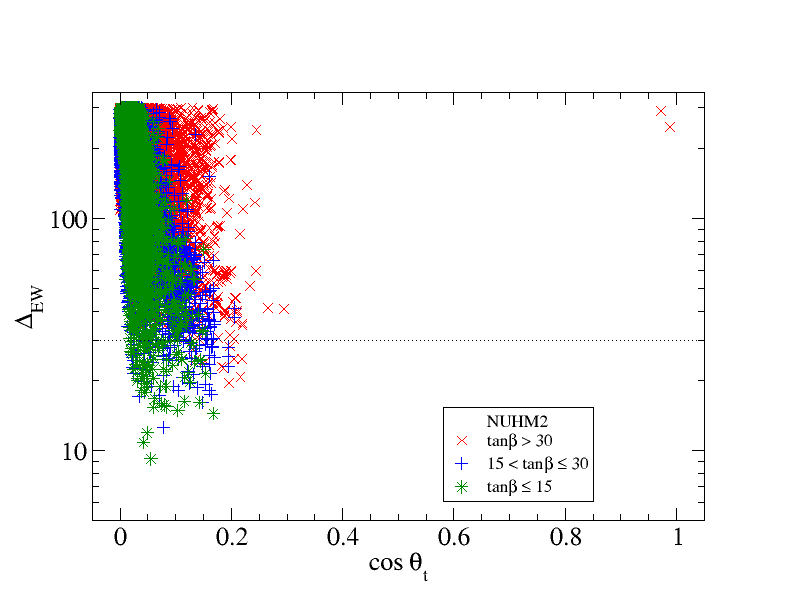}
\includegraphics[width=8cm,clip]{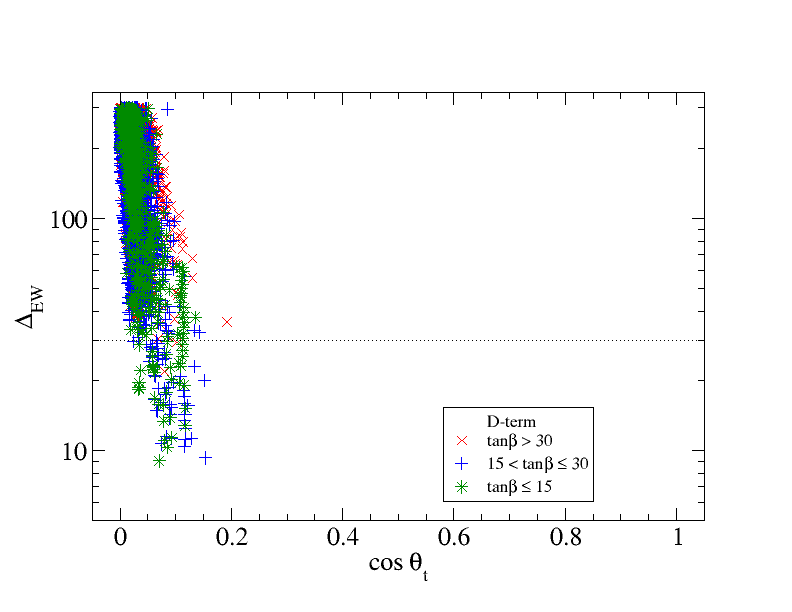}\\
\includegraphics[width=8cm,clip]{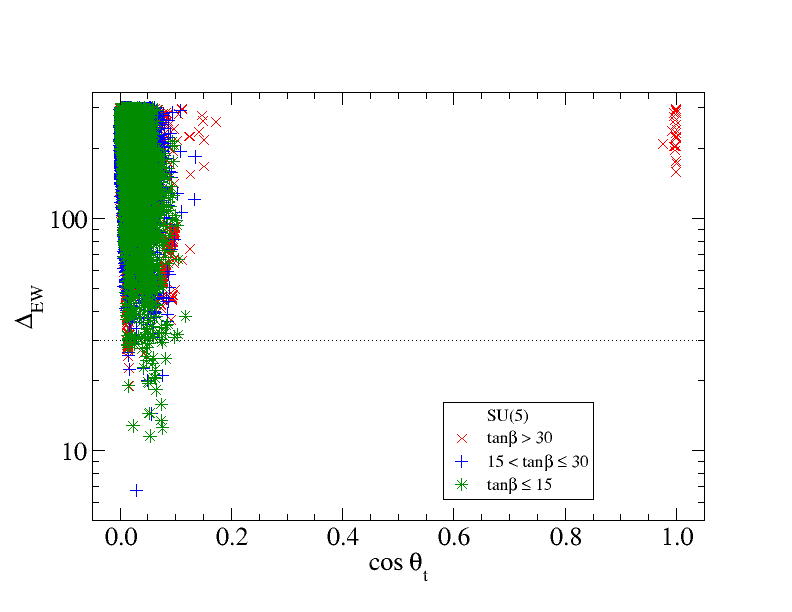}
\includegraphics[width=8cm,clip]{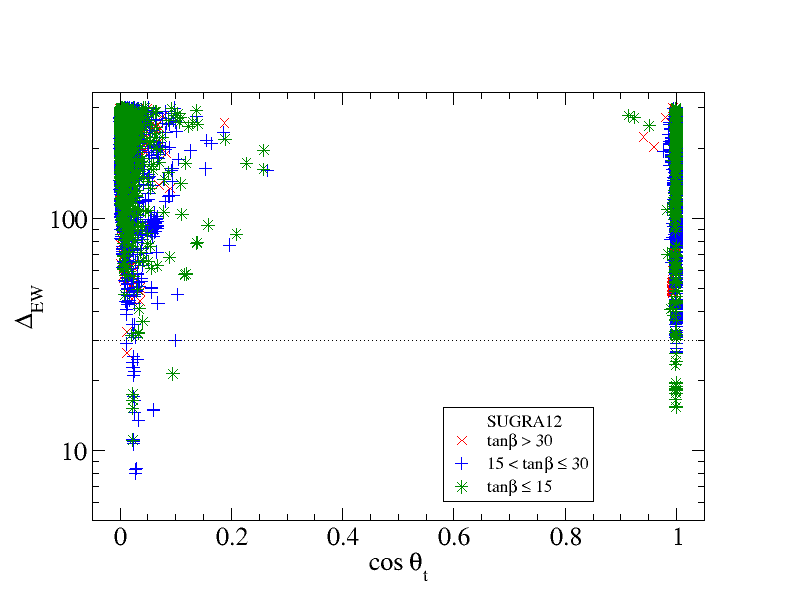}
\caption{Plot of $\Delta_{EW}$ vs. $\theta_{\tst}$ for {\it a}) the NUHM2 model, 
{\it b}) the D-term model, {\it c}) the $SU(5)$ model and {\it d}) 
the SUGRA12 model. 
\label{fig:thetat}}
\end{figure}

One aspect of the stop sector which may distinguish between the four models 
is listed in Fig. \ref{fig:thetat} where we plot the stop mixing angle
$\theta_t$ vs. $\Delta_{EW}$. Here we follow the notation of Ref. \cite{wss}
where $\tst_1=\cos\theta_t\tst_L-\sin\theta_t\tst_R$. Thus, $\cos\theta_t\sim 0$ 
leads to a $\tst_1$ which is mainly a right- state. From Fig. \ref{fig:thetat}
we see that for low $\Delta_{EW}<30$, then the NUHM2, DT and $SU(5)$ models
all require a mainly right- $\tst_1$. In constrast, the greater parameter freedom of the 
SUGRA12 model allows for low $\Delta_{EW}$ solutions with both 
left- and right- $\tst_1$ states.
If an $e^+e^-$ collider such as CLIC ($\sqrt{s}$ up to 3 TeV) is built with $\sqrt{s}>2m_{\tst_1}$, 
then the production cross sections for various beam polarizations will depend on 
the handedness of the stops being produced. Also, the left-stops
decay largely into charginos whilst the right-stops mainly decay only to neutralinos.
Such branching fraction measurements from an $e^+e^-$ collider could help to distinguish these cases.

\begin{figure}[tbp]
\includegraphics[width=8cm,clip]{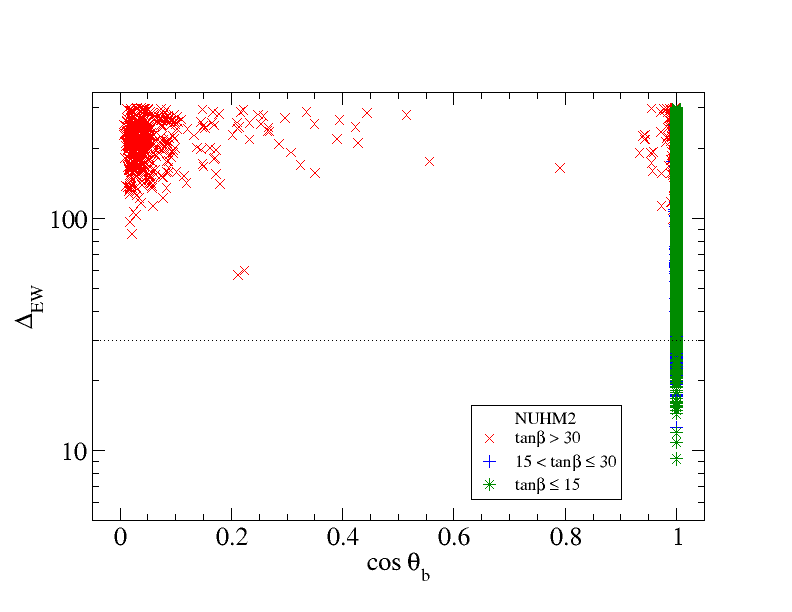}
\includegraphics[width=8cm,clip]{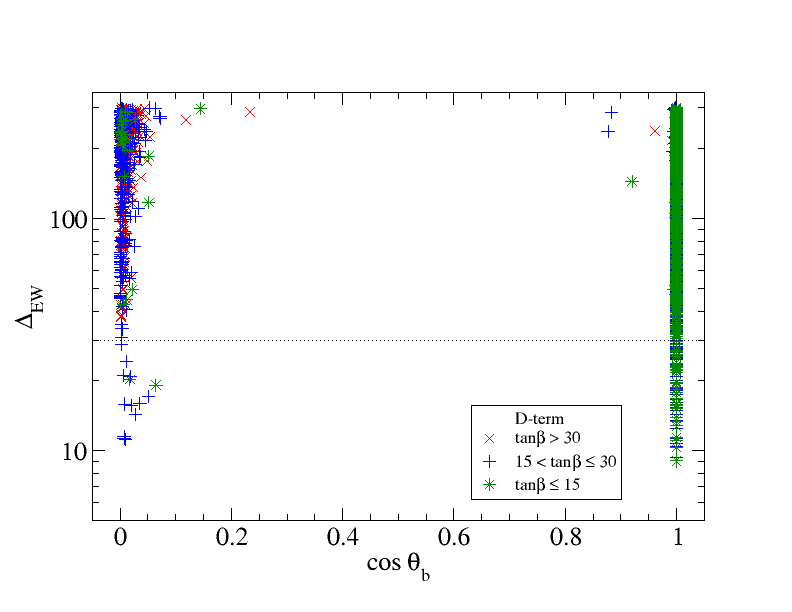}\\
\includegraphics[width=8cm,clip]{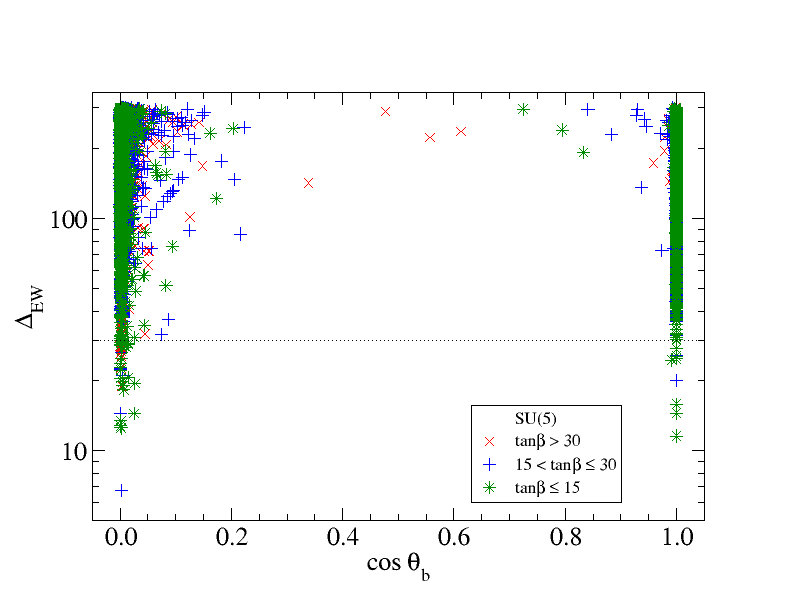}
\includegraphics[width=8cm,clip]{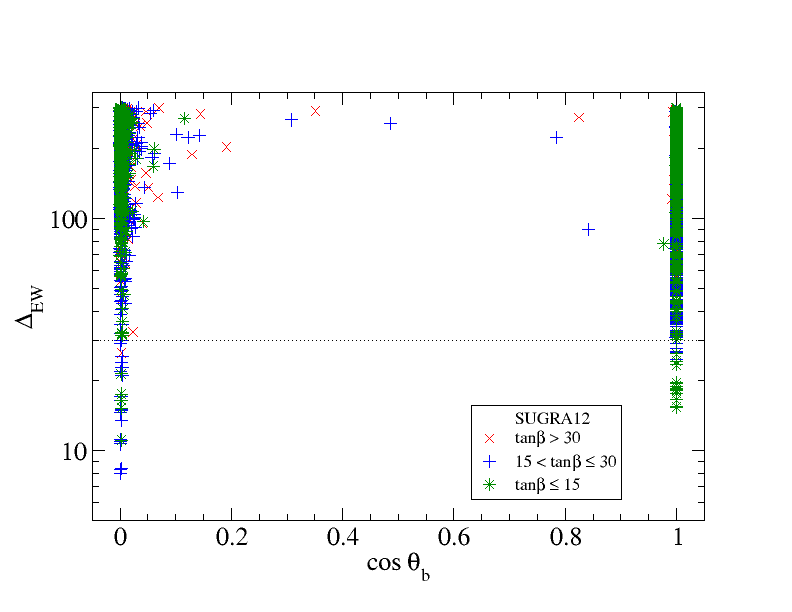}
\caption{Plot of $\Delta_{EW}$ vs. $\theta_{\tb}$ for {\it a}) the NUHM2 model, 
{\it b}) the D-term model, {\it c}) the $SU(5)$ model and {\it d}) 
the SUGRA12 model. 
\label{fig:thetab}}
\end{figure}

In the case of $\tb$-squarks, we list the corresponding mixing angle 
$\theta_b$ vs. $\Delta_{EW}$ for the four models in Fig. \ref{fig:thetab}. 
Here again, $\tb_1=\cos\theta_b \tb_L -\sin\theta_b\tb_R$. 
From the plots, we see that for natural solutions with $\Delta_{EW}<30$ in the NUHM2 model, 
then $\tb_1$ is expected to only occur as a left-squark. In the other three models, 
natural solutions exist where $\tb_1$ can occur as either left- or right- squarks. 
This can be understood in the NUHM2 model as a consequence of GUT scale universality: 
$m_{Q_3}=m_{D_3}$ where $m_{Q_3}$ is driven smaller than 
$m_{D_3}$ by the large top quark Yukawa coupling. 
For the other models where $m_{Q_3}$ may be greater than $m_{D_3}$ at
$Q=m_{GUT}$, then the lighter sbottom $\tb_1$ may be either left- or right-.

\begin{figure}[tbp]
\includegraphics[width=8cm,clip]{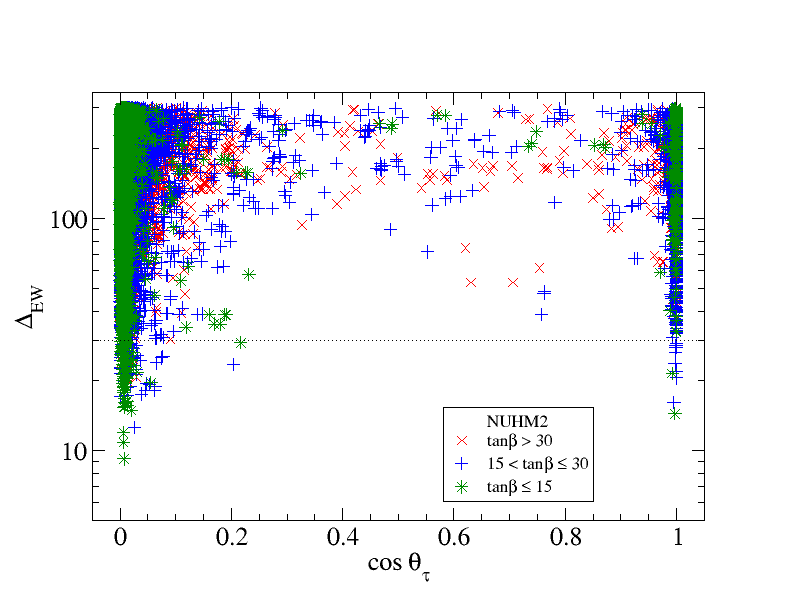}
\includegraphics[width=8cm,clip]{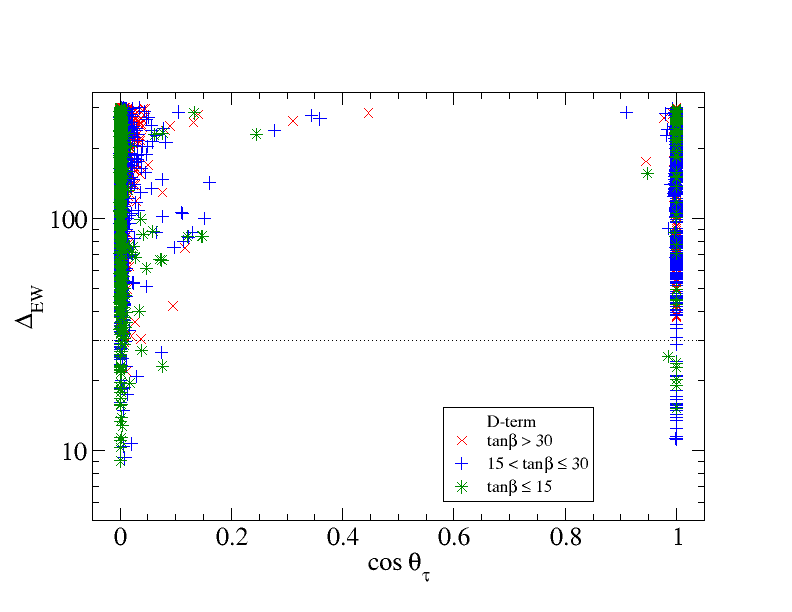}\\
\includegraphics[width=8cm,clip]{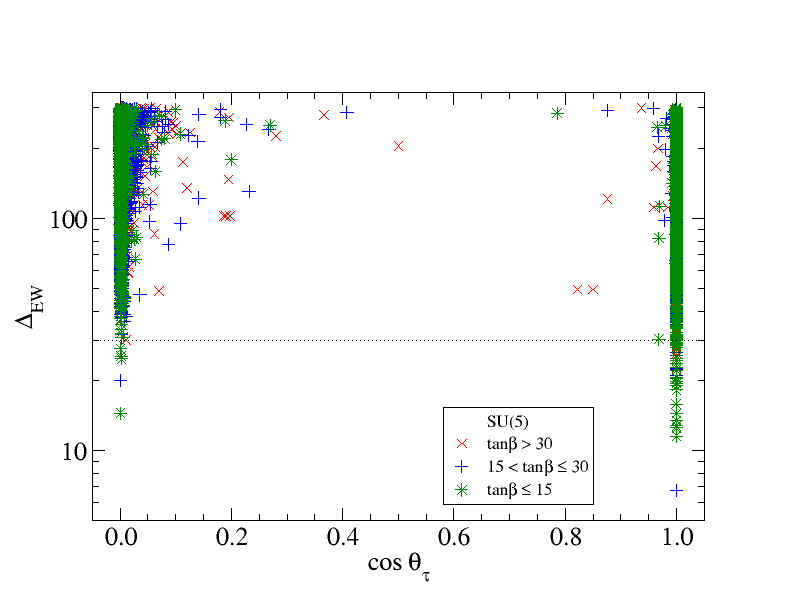}
\includegraphics[width=8cm,clip]{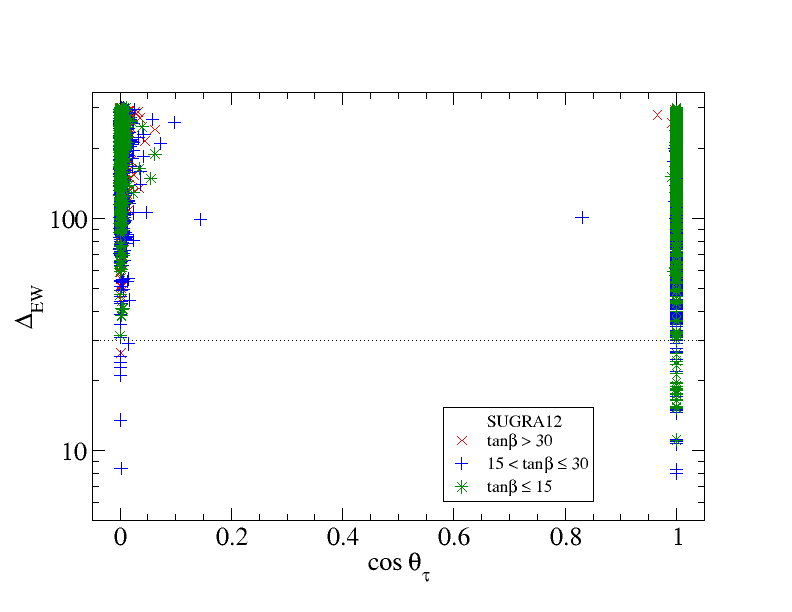}
\caption{Plot of $\Delta_{EW}$ vs. $\theta_{\ttau}$ for {\it a}) the NUHM2 model, 
{\it b}) the D-term model, {\it c}) the $SU(5)$ model and {\it d}) 
the SUGRA12 model. 
\label{fig:thetal}}
\end{figure}

In Fig. \ref{fig:thetal} we show the stau mixing angle $\cos\theta_\tau$ vs. 
$\Delta_{EW}$ where $\ttau_1=\cos\theta_\tau \ttau_L -\sin\theta_\tau\ttau_R$. 
In contrast to the stop and sbottom cases, 
we find that natural solutions with either right- or left staus 
can occur for all four models. 
Thus, meauring the ``handedness'' of the lighter staus is unlikely to distinguish between models.
Whereas in models like mSUGRA one always expects the lightest stau to be a right- state, 
in models with non-universality (at least in the Higgs sector) means that a large $S$ term contribution
($S=0$ in models with scalar mass universality) to RG running can reverse this situation 
and the lightest stau may in fact be a left- state.

\subsection{Squark and slepton masses}

To a very good approximation, the masses of first generation of sfermions are given by
\bea
m_{\tu_L}^2&=&
m_{Q_1}^2+m_u^2 +M_Z^2\cos 2\beta \left({1\over 2}-{2\over 3}\sin^2\theta_W\right)
\\
m_{\td_L}^2&=&m_{Q_1}^2+m_d^2 +M_Z^2\cos 2\beta \left(-{1\over 2}+
{1\over 3}\sin^2\theta_W\right)
\\
m_{\tu_R}^2&=&m_U^2+m_u^2 +M_Z^2\cos 2\beta \left({2\over 3}\sin^2\theta_W\right)
\\
m_{\td_R}^2&=&m_D^2+m_d^2 +M_Z^2\cos 2\beta \left(-{1\over 3}\sin^2\theta_W\right)
\\
m_{\te_L}^2&=&m_{L_1}^2+m_e^2 +M_Z^2\cos 2\beta \left(-{1\over 2}+\sin^2\theta_W\right)
\\
m_{\tnu_e}^2&=&m_{L_1}^2 +M_Z^2\cos 2\beta \left({1\over 2}\right)
\\
m_{\te_R}^2&=&m_E^2+m_e^2 +M_Z^2\cos 2\beta (-\sin^2\theta_W) ,
\eea
where the first terms on the right hand side of these expressions are the
weak scale soft SUSY breaking masses for the first generation of sfermions.
There are analogous expressions for second generation masses. 
It seems from a lack of signal from squark/slepton searches at LHC that
sfermion masses are likely in the multi-TeV region. In that case, the $D$-term contributions
to sfermion masses (those proportional to $M_Z^2$) are likely suppressed 
compared to the soft term contributions and hence the measured sfermion masses 
would very nearly provide the weak scale soft term masses.
The weak scale first/second generation soft terms have simpler RG running solutions 
so that a precise measurement of weak scale sfermion masses 
could yield the GUT scale soft terms\cite{zerwas}, especially if the gaugino masses
are measured. 
A knowledge of the GUT scale soft terms could then reveal whether or not the sfermions arrange 
themselves into GUT multiplets which would reflect a mass organization according to one
(or none) of the models considered.

\begin{figure}[tbp]
\includegraphics[width=8cm,clip]{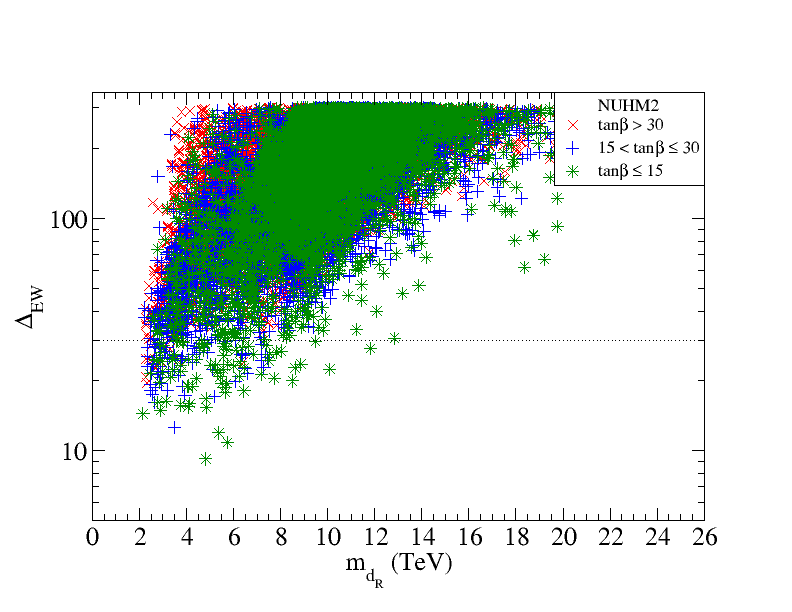}
\includegraphics[width=8cm,clip]{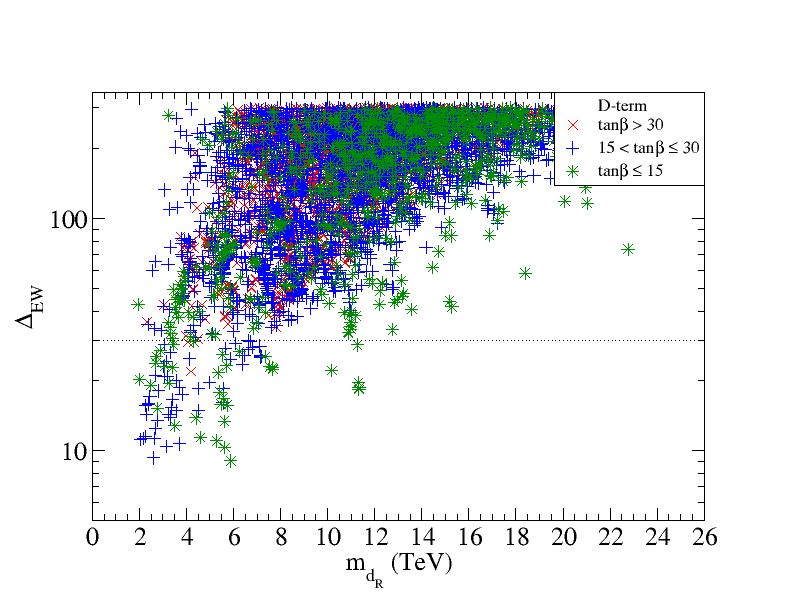}\\
\includegraphics[width=8cm,clip]{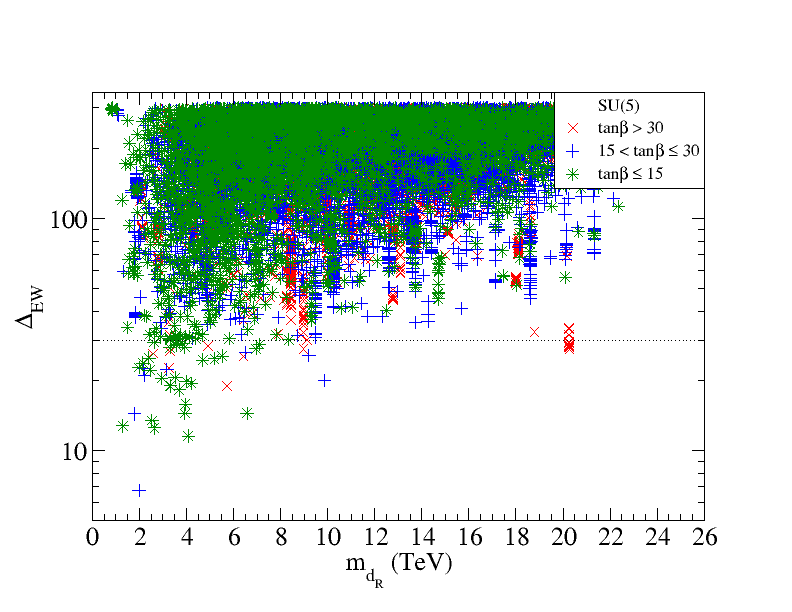}
\includegraphics[width=8cm,clip]{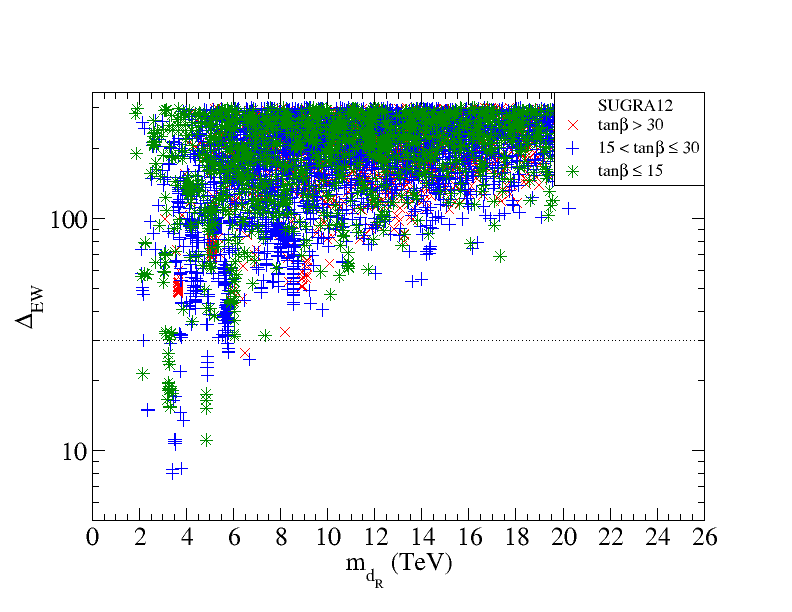}
\caption{Plot of $\Delta_{EW}$ vs. $m_{\td_R}$ for {\it a}) the NUHM2 model, 
{\it b}) the D-term model, {\it c}) the $SU(5)$ model and {\it d}) 
the SUGRA12 model. 
\label{fig:mdR}}
\end{figure}

In Fig. \ref{fig:mdR}, we show for example the $\td_R$ squark masses vs. $\Delta_{EW}$. 
While these squark masses may be as low as $\sim 2$ TeV for natural solutions, 
they can also range up to the vicinity of 10 TeV 
(and even up to 20 TeV for non-universal generations\cite{rns}). 
Thus, an $e^+e^-$ collider with $\sqrt{s}>2m(sfermion)$ would likely be required for
such squark mass determinations. Typically the $\sqrt{s}$ values needed would be beyond 
any sort of ILC projections and perhaps even beyond suggested CLIC energies. 
It is also possible such measurements could be made at a 50-100 TeV $pp$ collider
as suggested in Ref. \cite{hook}.
%
%
%

\subsection{Heavy Higgs masses}

\begin{figure}[tbp]
\includegraphics[width=8cm,clip]{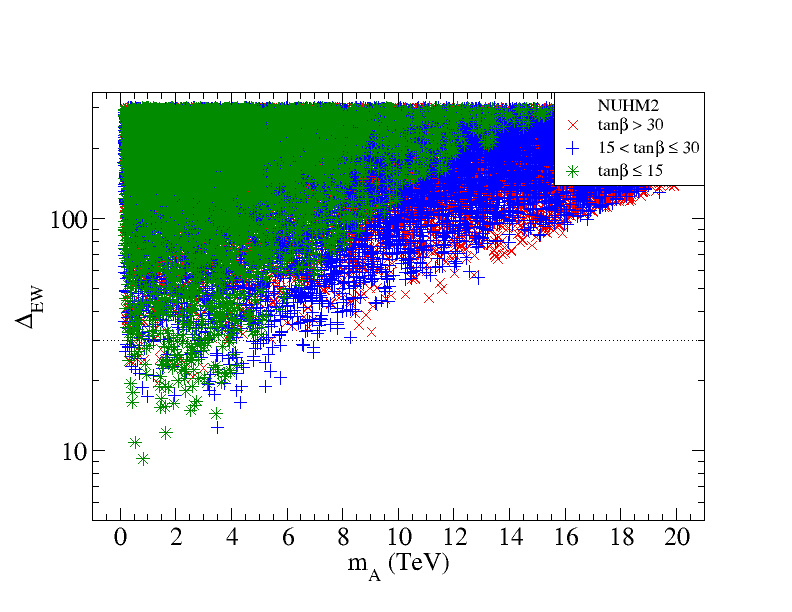}
\includegraphics[width=8cm,clip]{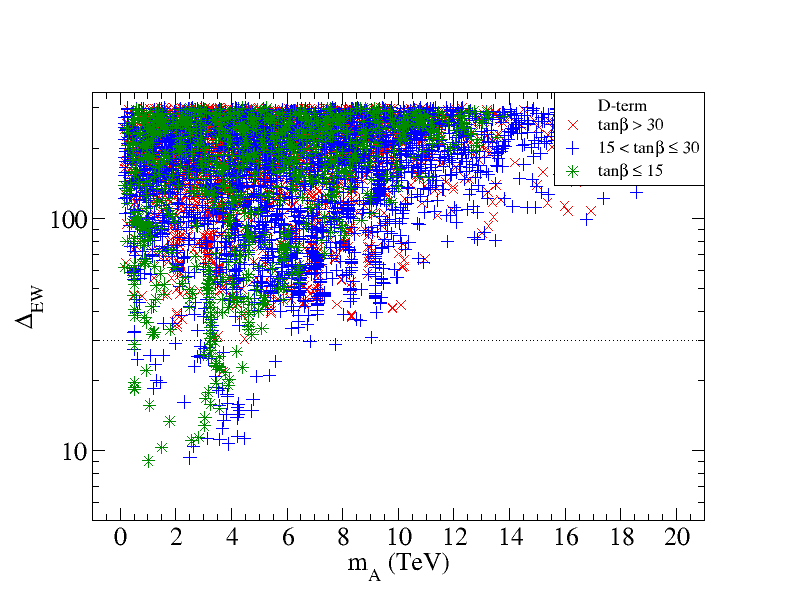}\\
\includegraphics[width=8cm,clip]{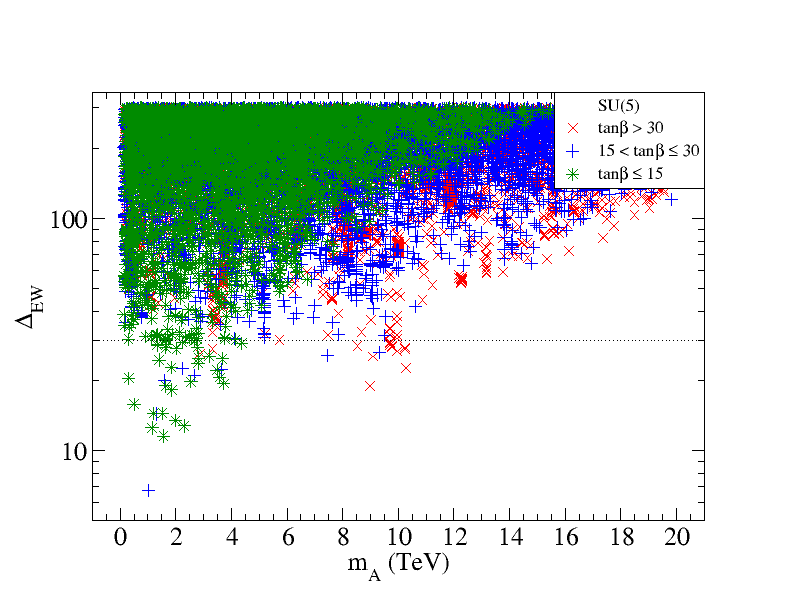}
\includegraphics[width=8cm,clip]{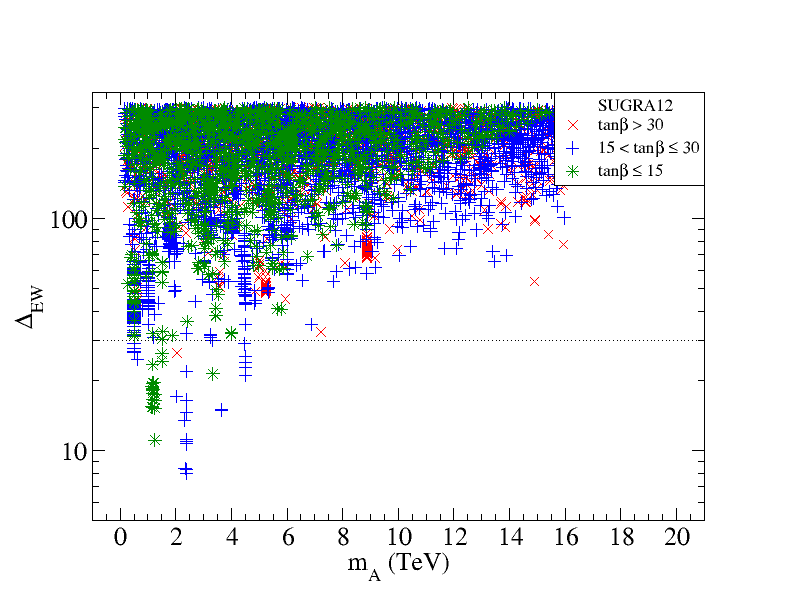}
\caption{Plot of $\Delta_{EW}$ vs. $m_A$ for {\it a}) the NUHM2 model, 
{\it b}) the D-term model, {\it c}) the $SU(5)$ model and {\it d}) 
the SUGRA12 model. 
\label{fig:mA}}
\end{figure}

Mass limits on heavy Higgs bosons have been shown previously for the NUHM2 model in
Ref. \cite{Hhiggs}. As confirmed in Fig. \ref{fig:mA}{\it a}), the value of $m_A$ is bounded by
about 8-10 TeV for this model. Similar mass bounds are found for the $DT$ model in frame {\it b})
and the $SU(5)$ model (frame {\it c}). For the SUGRA12 model in frame {\it d}), the mass bound appears
lower since now $D$-term contributions from first/second generation scalar masses come into play
in the $\Sigma_u^u$ terms in Eq. \ref{eq:mzs} and lead to unnaturalness for 
non-degenerate squarks and sleptons in the multi-TeV vicinity\cite{DTnat}.
Thus, the apparent tighter mass bound on $m_A$ in frame {\it d}) is likely due to
difficulty sampling at very high scalar masses.

\subsection{Four SUGRA GUT benchmark models}
\label{ssec:bm}

In Table \ref{tab:bm} we list four benchmark models, 
one for each model considered in the text. Each model
has $m_{1/2}=800$ GeV, $A_0=-5700$ GeV, $\tan\beta =10$ $\mu =150$ GeV 
and $m_A=3000$ GeV. The first case, NUHM2, has degenerate matter scalars
with mass $m_0=4$ TeV but with split Higgs mass soft terms. 
This model has low $\Delta_{EW}=23.7$ or 4\% EW fine-tuning. The gluino mass 
is $m_{\tg}=1972$ GeV which is somewhat above current limits from LHC13.
The higgsinos $\tw_1^\pm$ and $\tz_{1,2}$ are clustered around 150 GeV.
The $b-\tau$ Yukawa unification occurs at about 33\% level. 

The $DT$ model is listed next with input parameters $\mu$ and $m_A$ as listed.
These values determine $m_{H_u}^2(weak)$ and $m_{H_d}^2(weak)$ which are then run
up to $Q=m_{GUT}$ to determine the required $D$-term splitting. The matter 
scalars are split according to Eq. \ref{eq:dterms} leading to $m_{Q,U,E}=3597$ Gev and
$m_{D,L}=5019$ GeV. In spite of the different sfermion mass splitting, the 
value of $\Delta_{EW}$ remains at 23.4. The low energy spectrum of gluinos and 
higgsinos (and binos/winos) should ultimately be accessible to a combination
of LHC14 and ILC measurements. 
Since {\it all} four models have a similar spectrum of gauginos and higgsinos, 
they will all look rather similar to LHC14 and ILC.

Higher energy colliders such as CLIC ($\sqrt{s}\sim 3$ TeV) or a 100 TeV $pp$ collider 
$pp(100)$ will be required to 
distinguish the very massive sfermions. For the case of $DT$ model, 
measurements of $m_{\tu_L,\tu_R,\te_R}$ vs. $m_{\td_R,\te_L}$ would distinguish the
split rather degenerate matter scalars. A measurement of $m_A$ could help
determine $m_{H_d}^2(weak)$ which may then be run to $m_{GUT}$ to determine
$m_{H_d}(GUT)$. If knowledge of $m_{H_u}^2(GUT)$ can be extracted, 
then it might be possible to determine if the $D$-term splitting in the matter scalars 
is in accord with the Higgs soft mass splitting as in the 
DT model, or as in the $SU(5)$ model where $m_{Q,U,E}$ are split from $m_{D,L}$ in a manner quite 
different than the DT case. Note also that the $SU(5)$ model 
has a different pattern of stop-sbottom-stau mixing from the NUHM2 or $DT$ case
where now the $\tb_1$ is mainly a right-squark. 
The SUG12 model has a more arbitrary form of sfermion mass splitting. 
In this case, measurements that $m_{\td_R}\simeq m_{\te_R}$ and 
$m_{\tu_L}\sim m_{\tu_R}\sim m_{\te_L}$ would signal that the 
various matter sfermions do not live in GUT multiplets. In this latter case, 
there are incomplete cancellations of contributions to $\Delta_{EW}$
from the matter scalars\cite{DTnat} which may lift the calculated value 
of $\Delta_{EW}$ beyond what is otherwise expected.
\begin{table}\centering
\begin{tabular}{lcccc}
\hline
parameter & NUHM2 & D-term & SU(5) & SUG12 \\
\hline
$m_Q$      & 4000 & 3597 & 5000 & 5000 \\
$m_U$      & 4000 & 3597 & 5000 & 5000 \\
$m_E$      & 4000 & 3597 & 5000 & 3000 \\
$m_D$      & 4000 & 5019 & 3000 & 3000 \\
$m_L$      & 4000 & 5019 & 3000 & 5000 \\
$m_{H_u}$   & 4970 & 4648 & 5797 & 5468 \\
$m_{H_d}$   & 3043 & 3063 & 3022 & 3421 \\
\hline
$m_{\tg}$   & 1972.4 & 1965.1 & 1993.6 & 1989.9 \\
$m_{\tu_L}$ &  4250.3 & 3869.2 & 5194.8 & 5273.1 \\
$m_{\tu_R}$ &  4317.3 & 3928.1 & 5287.0 & 4949.6 \\
$m_{\td_R}$ & 4226.1  & 5220.2 & 3230.7 & 3485.9 \\
$m_{\te_L}$ &  4074.3 & 5074.1 & 3120.3 & 4819.7 \\
$m_{\te_R}$ &  3910.8 & 3517.3 & 4885.8 & 3596.4 \\
$m_{\tst_1}$&  1536.2 & 1060.0  & 2393.4 & 1798.9 \\
$m_{\tst_2}$&  3122.8 & 2758.1 & 3980.7 & 4103.6 \\
$m_{\tb_1}$ &  3146.4 & 2789.4 & 3163.6 & 3412.2 \\
$m_{\tb_2}$ &  4155.4 & 5147.7 & 3991.1 & 4137.6 \\
$m_{\ttau_1}$ & 3851.1 & 3445.3 & 3084.3 & 3528.8 \\
$m_{\ttau_2}$ & 4045.9 & 5044.8 & 4837.5 & 4795.1 \\
$m_{\tnu_{\tau}}$ & 4049.8 & 5054.1 & 3082.0 & 4797.5 \\
$m_{\tw_2}$ & 684.7 & 687.1 & 681.5 & 685.9 \\
$m_{\tw_1}$ & 154.8 & 154.4 & 155.9 & 155.9 \\
$m_{\tz_4}$ & 695.5 & 695.5 & 696.9 & 701.0 \\ 
$m_{\tz_3}$ & 359.7 & 359.8 & 360.0 & 360.8 \\ 
$m_{\tz_2}$ & 158.0 & 157.7 & 158.4 & 158.3 \\ 
$m_{\tz_1}$ & 142.0 & 141.7 & 142.5 & 142.4 \\ 
$m_h$      & 122.7 & 123.7 & 122.0 & 122.0 \\ 
\hline
$\Omega_{\tz_1}^{std}h^2$ & 0.008 & 0.008 & 0.008 & 0.008 \\
$BF(b\to s\gamma)\times 10^4$ & 3.0 & $2.9$ & 3.1 & 3.1 \\
$BF(B_s\to \mu^+\mu^-)\times 10^9$ & 3.8 & $3.8$ & 3.8 & 3.8 \\
$\sigma^{SI}(\tz_1 p)$ (pb) & $4.2\times 10^{-9}$ & $4.1\times 10^{-9}$ & 
$4.3\times 10^{-9}$ & $4.2\times 10^{-9}$ \\
$R_{b\tau}$ & 1.33 & 1.35 & 1.36 & 1.33 \\
$\Delta_{EW}$ & 23.7 & 23.4 & 54.0 & 37.6 \\
$\theta_t$ & 1.51 & 1.50 & 1.53 & 1.54 \\
$\theta_b$ & 0.0035 & 0.0012 & 1.57 & 1.56 \\
$\theta_\tau$ & 1.56 & 1.57 & 0.0015 & 1.57 \\
\hline
\end{tabular}
\caption{Input parameters and masses in GeV units for the four {\it
radiatively-driven natural SUSY} benchmark points from 1. the NUHM2 model
2. the D-term model, 3. the $SU(5)$ model and 4. the SUG12 model.
For all four cases, we take $m_{1/2}=800$ GeV, $A_0=-5700$ GeV, 
$\tan\beta =10$, $\mu=150$ GeV and $m_A=3000$ GeV. 
We also take $m_t=173.2$~GeV}.
\label{tab:bm}
\end{table}

\section{Conclusions:} 
\label{sec:conclude}

In this paper we have examined two topics: generalized focus point behavior of SUSY GUT models 
with radiatively-driven naturalness and a comparison of mass spectra expected from four different
SUSY GUT models. A crucial insight into naturalness was gleaned in Ref. \cite{fp} where it was 
demonstrated that for universal GUT scale boundary conditions on soft breaking scalar masses, 
large cancellations in the Higgs and squark contributions to the $Z$ boson mass allowed
for very heavy, TeV-scale third generation squarks whilst respecting naturalness. In our discussion of 
generalized focus-point behavior in Sec. \ref{sec:gfp}, we emphasized (as in Ref. \cite{seige}) that
in more fundamental SUSY theories (such as supergravity GUT theories) {\it all} the soft terms
are calculable as multiples of the gravitino mass $m_{3/2}$ (or $\Lambda$ in GMSB models) so that 
{\it all} the soft term contributions to $m_Z^2$ should be combined. In this situation, the
BG naturalness measure agrees with tree-level low electroweak fine-tuning as expressed by
the $\Delta_{EW}$ measure. We demonstrate for a hypothetical set of soft term relationships 
which link all the soft terms to $m_{3/2}$ that the weak scale value of $m_{H_u}^2$ is indeed
focussed to values $\sim m_Z^2$ for a wide range of gravitino mass values.

In the remainder of this paper we examined four scenarios expected from 
highly natural SUSY GUT models with gaugino mass unification but not scalar mass 
universality. The first task was to verify that all could generate low values of
$\Delta_{EW}\alt 30$. 
The next task was to examine how compatible $b-\tau$ Yukawa unification is with
electroweak naturalness and low $\mu$: we found them compatible to $R_{b\tau}\sim 1.2-1.5$
or 20-50\% $b-\tau$ Yukawa unification.
The third task was to examine the spectra from the four cases-- NUHM2,
$DT$, $SU(5)$ and SUGRA12 to examine if the models could be experimentally differentiable. 
In fact, all four models look rather alike for colliders like LHC14 and ILC.
For these cases, we expect the gluino mass to be bounded by about 4-5 TeV 
which may or may not be detectable at LHC. 
Also, a spectrum of light higgsinos with mass $\alt 200-300$ GeV are expected which
should be detectable at ILC. 
To differentiate the models, a very high energy hadron collider such as
a 100 TeV $pp$ machine will be needed for robust squark pair production or a very high
energy $e^+e^-$ machine will be needed for sfermion pair production. In such a case,
it may be possible to distinguish if the sfermions have nearby masses as expected in models like NUHM2
with matter scalar (but not Higgs) universality, 
or whether the spectrum is more spread out as expected in models with $D$-term splitting 
or where the sfermions come in independent ${\bf 10}$s and ${\bf 5^*}$s of
$SU(5)$. High energy $e^+e^-$ or $pp$ colliders may also be able to differentiate the decay
modes of third generation squarks to determine their ``handedness'', 
and determine if that agrees with expectations from various highly natural SUSY GUT models.

\section*{Acknowledgments}

This work was supported in part by the US Department of Energy, Office of High Energy Physics.

%

%
\end{document}